\documentclass[twocolumn,preprintnumbers,amsmath,amssymb,superscriptaddress,nofootinbib,longbibliography]{revtex4-2}

\usepackage[utf8]{inputenc}
\usepackage[english]{babel}

\usepackage{graphicx}

\usepackage{leftidx}
\usepackage{diffcoeff}
\difdef { f, fp, cp, c } {}
{
    outer-Ldelim = \left . ,
    outer-Rdelim = \right |,
    sub-nudge = 0 mu
}
\usepackage{amsfonts,amssymb,amsmath}
\usepackage{mathtools}
\usepackage{comment}

\usepackage{bm}
\usepackage{times}
\usepackage{bbm}
\usepackage{units}
\usepackage{mathrsfs}

\usepackage{array}
\usepackage{tabularx}
\usepackage{multirow}

\usepackage{hyperref}
\hypersetup{colorlinks=true,linktoc=all,linkcolor=blue,breaklinks=true,citecolor=blue,urlcolor=blue}

\usepackage[normalem]{ulem}
\usepackage{xspace}
\usepackage[dvipsnames]{xcolor}
\usepackage{dsfont}

\addto\captionsenglish{}

\AtBeginDocument{%
    \newwrite\bibnotes
    \def\bibnotesext{Notes.bib}
    \immediate\openout\bibnotes=\jobname\bibnotesext
    \immediate\write\bibnotes{@CONTROL{REVTEX42Control}}
    \immediate\write\bibnotes{@CONTROL{%
            apsrev42Control,author="08",editor="1",pages="1",title="0",year="1"}}
    \if@filesw
    \immediate\write\@auxout{\string\citation{apsrev42Control}}%
    \fi
}%

\newcommand*{\e}{\text{e}}
\newcommand*{\kB}{k_\text{B}}
\renewcommand*{\P}{\mathbf{P}}
\newcommand*{\bP}{\bar{\mathbf{P}}}
\newcommand*{\C}{\text{C}}
\newcommand*{\D}{\text{D}}
\renewcommand*{\H}{\text{H}}
\newcommand*{\W}{\text{W}}
\renewcommand*{\L}{\text{L}}
\newcommand*{\R}{\text{R}}
\newcommand*{\Jtrans}{J^Q_\text{trans}}
\newcommand*{\Jin}{J^Q_\text{in}}
\newcommand*{\JC}{J^Q_\C}
\newcommand*{\JH}{J^Q_\H}
\newcommand*{\JL}{J^Q_\L}
\newcommand*{\JR}{J^Q_\R}
\newcommand*{\GC}{\Gamma_\C}
\newcommand*{\GH}{\Gamma_\H}
\newcommand*{\GL}{\Gamma_\L}
\newcommand*{\GR}{\Gamma_\R}
\newcommand*{\UC}{U_\C}
\newcommand*{\UH}{U_\H}
\newcommand*{\T}{\bar{T}}
\newcommand*{\dT}{\delta T}
\newcommand*{\TL}{T_\mathrm{L}}
\newcommand*{\TR}{T_\mathrm{R}}
\newcommand*{\THot}{T_\mathrm{H}}
\newcommand*{\TCold}{T_\mathrm{C}}
\newcommand*{\eC}{\epsilon_\C}
\newcommand*{\eH}{\epsilon_\H}
\newcommand*{\eW}{\epsilon_\W}
\newcommand*{\Pcool}{\mathcal{P}_\text{cool}}
\newcommand*{\Scool}{{S}_{\Pcool\Pcool}}
\newcommand*{\Sin}{S_{\Jin\Jin}}
\newcommand*{\TUR}{\text{TUR}}
\newcommand*{\KUR}{\text{KUR}}
\newcommand*{\loc}{\text{loc}}
\newcommand{\K}{\mathcal{K}}
\newcommand{\I}{\text{(I)}}
\newcommand{\II}{\text{(II)}}

\newcommand*{\cC}{\mathcal{C}}
\newcommand*{\bcC}{\tilde{\mathcal{C}}}
\newcommand*{\CC}{\cC_\C}
\newcommand*{\bCC}{\bcC_\C}
\newcommand*{\CH}{\cC_\H}
\newcommand*{\bCH}{\bcC_\H}
\newcommand*{\CHC}{\cC_{\H\C}}
\newcommand*{\bCHC}{\bcC_{\H\C}}

\newcommand*{\GCop}{\GC^{0+}}
\newcommand*{\GCip}{\GC^{1+}}
\newcommand*{\GCom}{\GC^{0-}}
\newcommand*{\GCim}{\GC^{1-}}
\newcommand*{\GHop}{\GH^{0+}}
\newcommand*{\GHip}{\GH^{1+}}
\newcommand*{\GHom}{\GH^{0-}}
\newcommand*{\GHim}{\GH^{1-}}
\newcommand*{\GLoip}{\GL^{01+}}
\newcommand*{\GLiop}{\GL^{10+}}
\newcommand*{\GLoim}{\GL^{01-}}
\newcommand*{\GLiom}{\GL^{10-}}
\newcommand*{\GRoip}{\GR^{01+}}
\newcommand*{\GRiop}{\GR^{10+}}
\newcommand*{\GRoim}{\GR^{01-}}
\newcommand*{\GRiom}{\GR^{10-}}
\newcommand*{\GLoop}{\GL^{00+}}
\newcommand*{\GLoom}{\GL^{00-}}
\newcommand*{\GRoop}{\GR^{00+}}
\newcommand*{\GRoom}{\GR^{00-}}
\newcommand*{\g}{\gamma^5}
\newcommand*{\gC}{\gamma_\C^2}
\newcommand*{\gH}{\gamma_\H^2}
\newcommand*{\gCC}{\gamma_{\CC}^4}
\newcommand*{\gbCC}{\gamma_{\bCC}^4}
\newcommand*{\gCH}{\gamma_{\CH}^4}
\newcommand*{\gbCH}{\gamma_{\bCH}^4}
\newcommand*{\gCHC}{\gamma_{\CHC}^6}
\newcommand*{\gbCHC}{\gamma_{\bCHC}^6}
\newcommand*{\rC}{r_{\cC}}
\newcommand*{\rCC}{r_{\CC}}
\newcommand*{\rbCC}{r_{\bCC}}
\newcommand*{\rCH}{r_{\CH}}
\newcommand*{\rbCH}{r_{\bCH}}
\newcommand*{\rCHC}{r_{\CHC}}
\newcommand*{\rbCHC}{r_{\bCHC}}
\newcommand*{\cyc}{\text{cycle}}
\newcommand*{\tcyc}{\bar{t}_\cyc}
\newcommand*{\p}{\bar{p}}
\newcommand*{\cov}{\text{cov}}
\newcommand*{\var}{\text{var}}

\DeclarePairedDelimiter{\abs}{\lvert}{\rvert}
\DeclarePairedDelimiter{\mean}{\langle}{\rangle}
\DeclarePairedDelimiter{\cum}{\langle\!\langle}{\rangle\!\rangle}

\newcommand{\subfigref}[2]{\ref{#1}\hyperref[#1]{(#2)}}

\begin{document}

\title{Precision of an autonomous demon exploiting nonthermal resources and information}

\author{Juliette Monsel}
\affiliation{Department of Microtechnology and Nanoscience (MC2), Chalmers University of Technology, S-412 96 G\"oteborg, Sweden\looseness=-1}
\author{Matteo Acciai}
\affiliation{The Abdus Salam International Center for Theoretical Physics, Strada Costiera 11, 34151 Trieste, Italy}
\affiliation{Scuola Internazionale Superiore di Studi Avanzati, Via Bonomea 256, 34136, Trieste, Italy}
\affiliation{Department of Microtechnology and Nanoscience (MC2), Chalmers University of Technology, S-412 96 G\"oteborg, Sweden\looseness=-1}
\author{Didrik Palmqvist}
\affiliation{Department of Microtechnology and Nanoscience (MC2), Chalmers University of Technology, S-412 96 G\"oteborg, Sweden\looseness=-1}
\author{Nicolas Chiabrando}
\affiliation{Department of Microtechnology and Nanoscience (MC2), Chalmers University of Technology, S-412 96 G\"oteborg, Sweden\looseness=-1}
\author{Rafael S\'anchez}
\affiliation{Departamento de F\'isica Te\'orica de la Materia Condensada, Universidad Aut\'onoma de Madrid, 28049 Madrid, Spain\looseness=-1}
\affiliation{Condensed Matter Physics Center (IFIMAC), Universidad Aut\'onoma de Madrid, 28049 Madrid, Spain\looseness=-1}
\affiliation{Instituto Nicol\'as Cabrera (INC), Universidad Aut\'onoma de Madrid, 28049 Madrid, Spain\looseness=-1}
\author{Janine Splettstoesser}
\affiliation{Department of Microtechnology and Nanoscience (MC2), Chalmers University of Technology, S-412 96 G\"oteborg, Sweden\looseness=-1}

\date{\today}

\begin{abstract}
Quantum-dot systems serve as nanoscale heat engines exploiting thermal fluctuations to perform a useful task. Here, we investigate a multiterminal triple-dot system, operating as a refrigerator that extracts heat from a cold electronic contact. In contrast to standard heat engines, this system exploits a \textit{nonthermal} resource. This has the intriguing consequence that cooling can occur without extracting energy from the resource \textit{on average}---a seemingly demonic action--- while, however, requiring the resource to \textit{fluctuate}. Using full counting statistics and stochastic trajectories, we analyze the performance of the device in terms of the cooling-power precision, employing performance quantifiers motivated by the thermodynamic and kinetic uncertainty relations. We focus on two regimes with large output power, which are based on two operational principles: exploiting information on one hand and the nonthermal properties of the resource on the other. We show that these regimes significantly differ in precision. In particular, the regime exploiting the nonthermal properties of the resource can have cooling-power fluctuations that are suppressed with respect to the input fluctuations by an order of magnitude.  We also substantiate the interpretation of the two different working principles by analyzing cross-correlations between input and output heat currents and information flow.
\end{abstract}

\maketitle

\section{Introduction}

\begin{figure}[h!]
    \includegraphics[width=\linewidth]{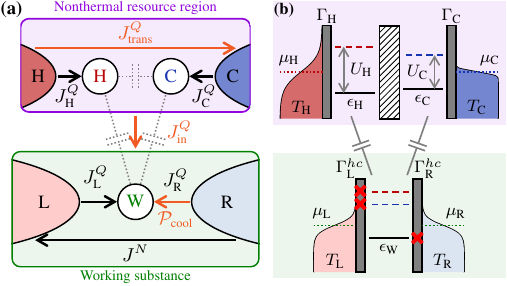}
    \caption{\label{fig:setup}
    (a) A resource region formed by two capacitively coupled quantum dots---each connected to one electronic reservoir, H or C---acts on a single quantum dot, W---connected to two reservoirs L and R---forming the working substance. A particle current $J^N$ can thereby be induced between contacts R and L, while heat currents $J_\alpha^Q$ flow out of (or into) each of the four reservoirs. The system works as a refrigerator when ${\cal P}_{\rm cool}=\JR>0$. (b) The reservoirs are kept at electrochemical potential $\mu_\alpha$ and temperature $T_\alpha$. Electrons tunnel between the reservoirs and the single level at energy $\eW+h\UH+c\UC$ in dot W or $\epsilon_j+wU_j$ in dots $j=$ H,C, where $h,c,w\in\{0,1\}$ are the occupations of dots H, C and W. Tunneling rates $\GH$ and $\GC$ are energy independent, while the tunneling rates, $\Gamma_\alpha^{hc}$ for dot W are energy dependent. We consider the ideal case where $\Gamma_\alpha^{hc}$ vanishes either when $h=c=0$ and $\alpha= \R$, or when either $h=1$ or $c=1$ and $\alpha=\L$ (as indicated by the red crosses).
    For a refrigerator configuration, we set $\TL>\TR$.
     }
\end{figure}

The thermodynamics of nanoscale and quantum systems has been at the focus of research in recent years~\cite{QTDBook2018} and attracts increasing attention~\cite{Campbell2026Jan}. A relevant class of example systems for solid-state implementations of thermodynamic machines are nanoelectronic devices such as quantum dot setups exploiting thermoelectric effects~\cite{Benenti2017Jun}. Several experiments have demonstrated steady-state heat-to-work conversion in two-terminal systems~\cite{Staring1993Apr,Dzurak1993Sep,Scheibner2007Jan,Svensson2012Mar,Josefsson2018Oct,Prete2019May,Volosheniuk2025Nov} and in multi-terminal systems where the coupling to the heat source is mediated by capacitively coupled quantum dots~\cite{Roche2015Apr,Hartmann2015Apr,Thierschmann2015Oct} or bosonic degrees of freedom~\cite{Dorsch2021Jan,Dorsch2021Dec,Haldar2024Feb}.
Importantly, in nanoscale systems, the resource of an engine does not necessarily have to be heat, it could also be information, as exemplified by the Maxwell demon~\cite{Koski2014Jul,Koski2015Dec,Chida2017May}.
The operation of the Maxwell demon or similarly of the Szilard engine is ``demonic" in the sense that it produces power while not exchanging energy with the resource, thereby seemingly violating the second law of thermodynamics~\cite{Whitney2023Apr,deOliveiraJunior2025Aug}.
Recently, it has been shown that also engines fed by a \textit{nonthermal} resource can be realized, having a seemingly demonic behavior, too. We associate a nonthermal resource with a distribution function that cannot be characterized by uniquely defined temperature and chemical potential. It has been shown that in such systems the \textit{average} energy exchange between resource and working substance can vanish, while the system produces useful work, again seemingly violating the second law of thermodynamics~\cite{Sanchez2019Nov,Ciliberto2020Nov,Deghi2020Jul,Hajiloo2020Oct,Lu2021Feb}. Nonthermal distributions have been recently used to achieve enhanced thermoelectric efficiency in interacting edge channels~\cite{Yamazaki2025Sep}. One way of realizing a system which incorporates features of both types of demons---the one exploiting information and the one exploiting the nonthermal resource---is to use capacitively coupled triple-dot systems~\cite{Whitney2016Jan,Sanchez2019Oct,Cao2025Jun,InfoTrajPaper}, see Fig.~\ref{fig:setup}.

In order to function, the nonthermal demonic systems require fluctuations in the input current, which hence only vanishes on average~\cite{Acciai2024Feb}. At the same time, it has been shown that---depending on the working principle, either information-based or heat-based---the precision of the output can differ significantly~\cite{InfoTrajPaper}. An in-depth analysis of the performance regarding the precision of this demonic quantum-dot engine is however missing.

In this paper, we fill this gap and broadly analyze the operation and performance of the three-dot refrigerator exploiting a nonthermal resource shown in Fig.~\ref{fig:setup} using full counting statistics~\cite{Landi2024Apr}. This gives us access to the fluctuations of particle- and energy currents, in particular of the cooling power, but also to information-current fluctuations and to the so-called activity of the device. In addition, we calculate cross-correlations between currents in different terminals and between different types of currents. Based on these calculations we achieve three main insights.
(1)~The cooling power fluctuations, as well as trade-off relations involving them, have different magnitudes depending on the operation principle of the device. In particular, we show how the device behaves with respect to performance quantifiers inspired by the thermodynamic and kinetic uncertainty relations~\cite{Barato2015Apr,DiTerlizzi2018Dec,Brandner2018Mar,Vo2022Sep,Brandner2025Jul}. (2)~The amount of output fluctuations, namely cooling power fluctuations, compared to the amount of input fluctuations, namely the (required) fluctuations in the incoming heat current, can be suppressed close to the maximum in cooling power, when the system exploits the nonthermal resource.  (3)~Finally, using cross-correlations between different heat currents and even information flows, we substantiate the interpretation of different working principles of the device and clearly identify why they lead to different performances concerning fluctuations. We support the interpretation of our results with insights from a complementary analysis of the system exploiting stochastic trajectories.

The remainder of this paper is organized as follows. We present the model and the observables of interest in Sec.~\ref{sec:approach}, where we also provide background on the full-counting statistics approach used to evaluate these observables. Details about stochastic trajectories, used as a comparison in this paper, are presented in an extensive appendix instead. We show results on the average cooling power and the exploited resources in Sec.~\ref{sec_res_coolingpower}, followed by the detailed analysis of the output fluctuations in terms of trade-off based performance quantifiers in Sec.~\ref{sec_tradeoff}, and the ratio between output and input fluctuations in Sec.~\ref{sec_res_noise}. We then analyze cross-correlations and the resulting Pearson correlation coefficients in Sec.~\ref{sec_res_cross}. Finally, in Sec.~\ref{sec_scatt}, we compare some of the above features with those occurring in a system where the energy exchange happens via particle exchange instead of the capacitive coupling of the main system, see Fig.~\ref{fig:setup}. Details about possible violations of tradeoff relations, about the explicit implementation of the full counting statistics, as well as about the stochastic trajectory analysis, can be found in the Appendices. We furthermore point to a Supplemental Material~\cite{Suppl}, where we present additional plots covering alternative observables and further parameter regimes.

\section{Theoretical approach}\label{sec:approach}

In this section, we introduce the model of the studied triple-dot system. The upper part forms a nonthermal resource region capacitively coupled to a working substance, where the resource is used to achieve cooling via energy filtering. While a generic nonthermal resource could serve as input, we here, for simplicity, take a mixture of two thermal reservoirs  at different temperatures to emulate the nonthermal resource.  We also introduce the observables of interest and show how to compute them using full counting statistics based on an extended master equation approach.

\subsection{Model}
\label{sec:model}
We consider a setup made of three capacitively coupled quantum dots, as depicted in Fig.~\ref{fig:setup}.
The lower part of this setting represents the working substance, which will here be operated as a refrigerator; it consists of the lower quantum dot W connected to two electronic contacts. The upper part constitutes the nonthermal resource region, consisting of the upper two dots H and C in contact with two further separate electronic contacts.
Such a system has previously been considered as a model for peculiar thermodynamic operations~\cite{Whitney2016Jan,Sanchez2019Oct,InfoTrajPaper,Cao2025Jun}, where a combination of information and nonthermal resource properties leads to a \textit{seemingly demonic} behavior.

For each dot, we consider a single level $\epsilon_i$, with $i=\H,\W,\C$, and assume that the intradot Coulomb interaction is larger than all other energy scales of the system. We therefore exclude double occupation~\cite{vanderWiel2002Dec} in each dot and furthermore neglect the spin of the electrons, as it could, for example, be achieved in the presence of a magnetic field.
The Hamiltonian describing the three quantum dots is then
\begin{equation}
    \hat{H}=\sum_{i=\H,\W,\C}\epsilon_i\hat{n}_i+\UH\hat{n}_\H\hat{n}_\W+\UC\hat{n}_\C\hat{n}_\W+U\hat{n}_\H\hat{n}_\C\ ,
\end{equation}
where $\hat{n}_i$ denotes the electron-number operator in dot $i$ and $\UH,\UC,U$ are interdot interactions.
Furthermore, we assume that the interaction energy between the two upper dots, H and C, is very large, $U \to \infty$, so that double occupation of the resource dots is excluded. This simplification is convenient to separate the effects of their respective fluctuations on the system dynamics, but the observed features persist even for finite $U$, see Ref.~\cite{Whitney2016Jan}.

Each dot of the resource region is tunnel-coupled to a single terminal, that is a large fermionic reservoir, characterized by a temperature $T_\alpha$ and chemical potential $\mu_\alpha$---these are indicated by $\alpha = \C,\H$. We choose to have $\TCold < \THot$ and therefore call the corresponding terminals the cold and hot resource reservoirs. Note again that the nonthermal resource created by this upper subsystem does not rely on being constructed from thermal reservoirs. However, here we use this scheme for convenience.
The dot W, in the lower, working-substance part of the setup, is tunnel-coupled to two terminals---these are indicated by $\alpha = \L,\R$, characterized by temperatures $\TL$, $\TR$, and chemical potentials $\mu_\L$, $\mu_\R$, respectively. In the following, we denote the average temperature of the working substance reservoirs by $\bar{T} = (\TL + \TR)/2$. Without loss of generality, we set the energy references such that $\mu_\L = \mu_\C = \mu_\H = 0$. Furthermore, since we are interested in operating the working substance as a refrigerator, we consider a temperature difference $\delta T=\TL-\TR\ge0$, aiming at cooling down the right reservoir, and set $\mu_\R\equiv 0$.
With these parameters, the four reservoirs are characterized by Fermi functions $f_\alpha(E) = 1/(1 + \e^{(E - \mu_\alpha)/k_\mathrm{B}T_\alpha})$.

The coupling between the quantum dots and the electronic reservoirs is characterized by the tunnel-coupling strengths $\Gamma_\alpha$. While we assume the tunnel couplings $\Gamma_\C,\Gamma_\H$ of the resource region to be constant, those of the working substance depend on the energy at which the electrons tunnel. Since this energy is fully determined by the occupations $h,c=0,1$ of the resource dots H and C, we denote these tunnel rates $\Gamma_{\L/\R}^{hc}$.
In this work, we consider the weak-coupling (sequential-tunneling) regime, i.e., $\hbar\Gamma_\alpha \ll \kB T_\alpha$, and, in the following, we use the convention $\hbar, \kB \equiv 1$.

\subsection{Observables of interest}\label{sec:observables}

We are interested in the operation of the engine as a steady-state refrigerator and aim to analyze the cooling power, the efficiency, and, in particular, the precision. The observables of interest are hence the steady-state heat currents
\begin{equation}
    J_\alpha^Q=J^E_\alpha-\mu_\alpha J^N_\alpha
\end{equation}
defined by the energy currents, $J^E_\alpha$, and the particle currents, $J^N_\alpha$. Since we have here chosen all electrochemical potentials to be zero, the heat currents always equal the energy currents. We define currents as being positive when flowing \textit{out of} the reservoir. In the stationary regime of interest here, for the resource particle currents, we always have $J_\C^N=J_\H^N=0$ since each resource dot is coupled to only one reservoir. From charge and energy conservation, we additionally have $J^N\equiv J_\R^N= - J_\L^N$ and $\sum_\alpha J_\alpha^E=0$.

In particular three (combinations of) heat currents are of interest here. This is on one hand, the cooling power, which is the \textit{task} of the engine
\begin{equation}
    0\leq \Pcool=\JR\ .
\end{equation}
On the other hand, the heat currents of the resource can be combined in the meaningful combinations
\begin{eqnarray}
    \Jin & = & \JC+\JH\ ,\label{eq:Jin}\\
    \Jtrans & = & \frac{1}{2}\left(\JH-\JC\right)\ .\label{eq:Jtrans}
\end{eqnarray}
The heat current injected from the resource region into the working substance is given by $\Jin$. Without constraints on the injected current, the setup would work as an absorption refrigerator~\cite{Sanchez2011Feb,Levy2012Feb,Venturelli2013Jun,Erdman2018Jul}. The action of the autonomous demon is to achieve refrigeration under the condition that no heat is injected into the working substance~\cite{InfoTrajPaper,Picatoste2024Aug}, i.e., we refer to $\Jin\equiv0$ as the demon condition.  We furthermore define the heat current in the resource region as $\Jtrans$.

Besides heat, information is an important resource for refrigeration, in particular when the system operates under the demon condition~\cite{Strasberg2013Jan,InfoTrajPaper}. The mutual information between the two subsystems is given by the difference between the von Neumann entropy, $\mathcal{S}$,  of the total system compared to the ones of the separate subsystems $I=\mathcal{S}_\mathrm{ws}+\mathcal{S}_\mathrm{res}-\mathcal{S}$. We quantify the change in mutual information with the information current $J^I$ from the working substance, corresponding to the rate at which the resource region acquires information about the working substance \cite{Ptaszynski2019Apr}.

We are furthermore interested in fluctuations, $S_{J^Q_\alpha J^Q_\alpha}$, in the heat currents as well as in correlations between them, $S_{J^Q_\alpha J^Q_{\gamma}}$. Also fluctuations in particle and information currents, $S_{J^NJ^N}$ and $S_{J^IJ^I}$, will be considered. We obtain them from full counting statistics as explained in the following section. The fluctuations and correlations yield information about the operation of the engine, but in particular also quantify the precision of the outcome.

To set bounds on trade-offs between cooling power, its efficiency, and precision, we resort to the recently studied thermodynamic~\cite{Barato2015Apr,Gingrich2016Mar,Horowitz2020Jan} and kinetic uncertainty relations~\cite{DiTerlizzi2018Dec,Vo2022Sep}.
First, we consider the performance quantifier deriving from the thermodynamic uncertainty relation (TUR)
\begin{equation}\label{Q_TUR}
    X_\TUR = \frac{2\Pcool^2}{\Scool\dot{\Sigma}} \le 1\ ,
\end{equation}
containing the total entropy production rate $\dot{\Sigma}$. The latter can here be obtained by sums over heat currents weighted by temperatures of the separate reservoirs \cite{Ptaszynski2019Apr}, $\dot{\Sigma}=-\sum_\alpha J_\alpha^Q/T_\alpha$\footnote{The entropy production rate is given more generally by $\dot{\Sigma}=\diff{\mathcal{S}}{t}-\sum_\alpha J_\alpha^Q/T_\alpha$, but since we are considering the steady state, the von Neumann entropy of the system remains constant, $\diff{\mathcal{S}}{t} = 0$.}. In addition, we consider a performance quantifier deriving from the kinetic uncertainty relation (KUR)
\begin{equation}\label{Q_KUR}
    X_\KUR = \frac{\Pcool^2}{\Scool \K} \le 1\ ,
\end{equation}
which contains the so-called activity $\K$~\cite{DiTerlizzi2018Dec, Landi2024Apr}.
We will also analyze a local version of the KUR,
\begin{equation}\label{Q_KUR_loc}
    X_\KUR^\loc = \frac{\Pcool^2}{\Scool S_{J^N\!J^N}}\ ,
\end{equation}
introduced in Ref.~\cite{Palmqvist2025Oct} for noninteracting systems within scattering theory, where the local activity is associated to particle-current fluctuations $S_{J^N\!J^N}$. In coherent quantum transport Eq.~\eqref{Q_KUR_loc} was found to be bounded by one, either close to equilibrium or in the tunneling regime, where the local activity of a reservoir is well approximated by the particle current noise.
Since the triple-dot device in this work behaves according to a classical Markov process, both bounds in Eqs.~\eqref{Q_TUR} and~\eqref{Q_KUR} hold. Instead,
due to the capacitive coupling between the resource region and the working substance (out of the initial scope of Ref.~\cite{Palmqvist2025Oct}), it is no longer guaranteed that $X_\KUR^\loc \le 1$. Indeed, we find parameter regimes where this bound is broken, as shown in Appendix \ref{app:breaking loc KUR}. However, the $X_\KUR^\loc$ quantifier has the advantage of being fully expressed in terms of transport observables. Similar uncertainty relations have recently been discussed, such as the thermokinetic~\cite{Vo2022Sep,VanVu2025Mar,Palmqvist2025Jul} or clock uncertainty relation~\cite{Prech2025Sep}. Their detailed analysis goes beyond the scope of the present paper, but we show some results for completeness in the Supplemental Material \cite{Suppl}.

In order to investigate the type of demon-like operation of the device, we furthermore compute the so-called Pearson correlation coefficient~\cite{Freitas2021Mar}\footnote{This definition is similar but not exactly equal to the Pearson correlation coefficient from Ref.~\cite{Freitas2021Mar}, which was defined in terms of stochastic trajectories, see the discussion in Appendix \ref{app:Pearson}.}
\begin{equation}\label{r}
    \varrho_{JJ'} = \frac{S_{JJ'}}{\sqrt{S_{JJ}S_{J'J'}}}\ ,
\end{equation}
for correlations between heat currents into different terminals and combinations between heat currents, such as $\Jin$ and $\Jtrans$ introduced in Eqs.~\eqref{eq:Jin} and \eqref{eq:Jtrans}.

\subsection{Evaluation of observables from full counting statistics}\label{sec:model:FCS}

We use steady-state full counting statistics~\cite{Bagrets2003Feb,Flindt2008Apr,Utsumi2015Oct} to compute the particle, energy, and information currents and activity, as well as current fluctuations~\cite{Andrieux2009Apr}. Counting statistics of charge can even be accessed experimentally: this is done by time-resolved measurement of the current in capacitively coupled quantum point contacts~\cite{fujisawa:2006,kung_irreversibility_2012,Hofmann2016Jan,Hofmann2017Mar,Chida2022Oct} or single-electron transistors~\cite{Saira2012Oct,Koski2013Oct,Koski2014Jul,Maillet2019Apr}.  In Coulomb blockaded systems such as the one considered here, one can associate a precise energy to every particle exchange, and it is therefore in principle possible to access the heat counting statistics with a charge detector as well~\cite{Sanchez2012Dec}, as recently realized experimentally~\cite{Chida2025Dec}.

Average currents and their fluctuations are given by the cumulants (indicated by $\cum{\bullet}$) of the respective transport quantity. Particle, information, and energy currents, as well as the activity, are found as the time derivatives of the average values (first-order cumulants)
\begin{align}\label{J_FCS}
    J^N&=\diff{\cum{N}}{t}\ , &   J^Q_\alpha&=J^E_\alpha=\diff{\cum{E_\alpha}}{t}\ , \\\nonumber
    J^I&=\diff{\cum{I}}{t}\ ,& \K &\equiv J^A=\diff{\cum{A}}{t}\ .
\end{align}
We have denoted $N$ the number of particles transferred into dot W from reservoir R, $E_\alpha$ the energy transferred from reservoir $\alpha = $ C, H, L, R, into the corresponding dot, $I$ the information on the working substance acquired by the resource region, and $A$ the total number of hopping events in a given measuring time.
The second-order cumulants yield
\begin{align}\label{S_FCS}
    S_{J^N J^N} &=  \diff{\cum{N^2}}{t}\ ,&  S_{J^I J^I} &=  \diff{\cum{I^2}}{t}\ ,\\\nonumber
    S_{J_{\alpha^{}}^Q J_{\gamma}^Q}   &=   \diff{\cum{E_{\alpha} E_{\gamma}}}{t}\ ,& S_{J^I J_{\alpha}^Q}   &=   \diff{\cum{I E_{\alpha}}}{t}\ ,
\end{align}
namely, the fluctuations, correlations between currents in different contacts, and correlations between different currents.

To compute these quantities, the cumulant generating functions need to be found. We start by defining counting fields for the particles flowing out of the right reservoir, $\xi^N$, for the energy flowing out of reservoir $\alpha$, $\xi^E_\alpha$, for the total amount of hopping events, $\xi^A$, and for the information acquired by the resource, $\xi^I$.
In the regime of weak coupling,  where coherences between dot states do not play a role for the transport quantities\footnote{Coherences do not couple to probabilities in the weak-coupling regime as long as there is no hopping between different quantum dots and no spin precession occurring in the dynamics.}, one can then set up an extended master equation including the counting fields~\cite{Sanchez2013Dec,Schaller2014Jan, InfoTrajPaper}
\begin{equation}
\label{eq:extmastereq}
    \difcp{\P}{t}(t, \bm{\xi}) = W(\bm{\xi})  \P(t, \bm{\xi})\ ,
\end{equation}
with $\bm{\xi} = (\xi^N, \xi^E_\C, \xi^E_\H, \xi^E_\L, \xi^E_\R, \xi^A, \xi^I)$. Since we here consider the limits of strong capacitive interactions on each dot and between dots H and C, the vector $\P(t, \bm{\xi})$ contains the elements $\P = (p_{000}, p_{001}, p_{010}, p_{011}, p_{100}, p_{110})^\text{T}$, where the subindex ${h w c}$ denotes the states where the dots H, W and C contain respectively $h,w,c=0,1$ electrons.
The kernel $W(\bm{\xi})$ is hence a $6\times6$ matrix which can be decomposed into contributions from each terminal, $W(\bm{\xi}) = W_\C(\xi^E_\C,\xi^A,\xi^I) + W_\H(\xi^E_\H,\xi^A,\xi^I) + W_\L(\xi^E_\L,\xi^A) + W_\R(\xi^N, \xi^E_\R,\xi^A)$. The full expressions of the kernel elements and their dependence on the counting fields are given in Appendix~\ref{app:FCS kernel}.

When setting all counting fields to zero, Eq.~\eqref{eq:extmastereq} yields a standard master equation for the occupation probabilities of the triple-dot system, $\partial_t\mathbf{P}=W\mathbf{P}$~\cite{Beenakker1992Oct}. We are interested in the steady state of the setup, $\bP$, such that $W\bP = 0$, see Ref.~\cite{InfoTrajPaper} for an analytic solution of $\bP$.

Instead, when keeping all counting fields finite, we obtain the cumulant-generating function $\mathcal{F}(\bm{\xi}) = \ln[\mathcal{G}(t, \bm{\xi})]$, where $\mathcal{G}(t, \bm{\xi}) = \sum_{hwc}[ \P(t, \bm{\xi})]_{hwc}$ is the moment-generating function. Then, the combined cumulants of the transferred particle number, energy, hopping events, and information are obtained by differentiating
   \begin{equation}
\begin{split}
\cum{\Xi^kE_\alpha^{k'}} &= (-i)^{k+k'}\difcp[k,k']{\mathcal{F}(t, \bm{\xi})}{\xi^\Xi,\xi^E_\alpha}[\bm{\xi}=0]\ ,\\
\cum{E_\alpha^{k}E_\gamma^{k'}} &= (-i)^{k+k'}\difcp[k,k']{\mathcal{F}(t, \bm{\xi})}{\xi^E_\alpha,\xi^E_\gamma}[\bm{\xi}=0]\ ,
\end{split}
\end{equation}
   with $\Xi = N, A, I$.
We are interested in the long-time limit in order to study the steady state. In this case, the cumulant-generating function can be approximated by $\mathcal{F}(t, \bm{\xi}) \simeq \lambda(\bm{\xi})t$ \cite{Schaller2014Jan}, where $\lambda(\bm{\xi})$ is the eigenvalue of $W(\bm{\xi})$ such that $\lambda(\bm{\xi}) \to 0$, when $\bm{\xi}\to 0$, while all other eigenvalues have negative real parts in this limit. This is, of course, valid only if the system dynamics have a unique steady state, $\bP$, which is the case here.

One can then obtain the currents and second-order cumulants (auto- and cross-correlations) by taking derivatives of $\lambda(\bm{\xi})$ with respect to the corresponding counting fields.
However, the diagonalization of the kernel quickly becomes cumbersome with increasing system complexity. Here, we are only interested in the first two cumulants (mean currents and their correlations), in which case a recursive method is more appropriate and allows us to treat the problem analytically~\cite{Lenstra1982Dec,Sanchez2007Apr}. We start by taking the Laplace transform of the moment-generating function in the long-time limit \cite{Sanchez2013Dec, Schaller2014Jan}, \begin{equation}\label{G(z)}
    \mathcal{G}(z, \bm{\xi}) = \sum_{hwc}\left[\left(z\mathds{1} - W(\bm{\xi})\right)^{-1}\P_0\right]_{hwc}\ ,
\end{equation}
where $\P_0$ is the initial state. The initial state plays no role in the stationary regime, so we can choose it as simple as possible to ease the calculation. We therefore take $\P_0 = \bP$. The eigenvalue $\lambda(\bm{\xi})$ we want to compute is the pole $z_0$ of $\mathcal{G}$ close to $\bm{\xi} = 0$. We therefore take the Taylor expansion of $z_0$ around $\bm{\xi} = 0$,
\begin{align}\label{z0}
\begin{split}
    z_0 = \sum_{n, \{m_\alpha\}_\alpha, k,\ell \in\mathbb{N}} c_{n,m_\C ,m_\H,m_\L,m_\R,k,\ell} \\
    \times\frac{(i\xi^N)^n}{n!}\left[\prod_\alpha \frac{ (i\xi^E_\alpha)^{m_\alpha}}{m_\alpha!}\right]
    \frac{(i\xi^A)^k}{k!}
    \frac{(i\xi^I)^\ell}{\ell!}\ ,
    \end{split}
\end{align}
in Eq.~\eqref{G(z)}, do the inverse Laplace transform and then determine the coefficients $c_{n,m_\C ,m_\H,m_\L,m_\R,k,\ell}$ by proceeding order by order. The coefficients $c_{n,m_\C,m_\H,m_\L,m_\R,k,\ell}$ directly correspond to the currents, fluctuations, and cross-correlations. Explicitly, we have $J^N = c_{1,0,0,0,0,0,0}$, $\JC = c_{0, 1, 0, 0, 0,0,0}$, $\K = c_{0,0,0,0,0,1,0}$, $J^I = c_{0,0,0,0,0,0,1}$, $S_{J^N J^N} = c_{2, 0, 0, 0, 0,0,0}$,  $S_{\JC\JC} = c_{0, 2, 0, 0, 0,0,0}$,  $S_{\JC\JH} = c_{0, 1, 1, 0, 0,0,0}$, and so on.

\section{Fluctuations and correlations of a demonic refrigerator}

The task of the device depicted in Fig.~\ref{fig:setup} is to extract heat from the colder of the two contacts in the working substance, namely contact R, while dumping heat into the hotter contact L. This is done by exploiting the resource provided by the upper part of the device. While at the trajectory level (see also the sketch in Fig.~\ref{fig:cycles}) energy is exchanged between resource and working substance, this energy transfer averages out under the demon condition, while entropy production in the resource region and information flow between the two subsystems persist.

The operation of the nonthermal engine depicted in Fig.~\ref{fig:setup} depends on a large set of parameters. Here, we analyze in detail different types of fluctuations and correlations of two complementary settings. In particular, we focus on two scenarios of large cooling power, for optimal level positions of the quantum dots C and W, as previously identified in Ref.~\cite{InfoTrajPaper}, while the level position of dot H is tuned to fulfill the demon condition $\Jin = 0$. We will introduce these scenarios in the next section, Sec.~\ref{sec_res_coolingpower}.
Like in Ref.~\cite{InfoTrajPaper}, we assume in the following an ideal tunnel-rate asymmetry in the working substance\footnote{Note that this is not a strict requirement for the refrigerator to work, as discussed in Appendix \ref{app:tunneling asym}.}---namely $\GR^{00} = \GL^{01} = \GL^{10} = 0$, as indicated by the red crosses in Fig.~\subfigref{fig:setup}{b}. Also, we always choose $\GR,\GL\ll\GH,\GC$, corresponding to an ideal demon, which measures and reacts rapidly.

In the following, we analyze the precision of the cooling power, the relation between input and output fluctuations, and we further substantiate predictions from Ref.~\cite{InfoTrajPaper}, where the two different regimes of operation were identified as Maxwell-demon-like~\cite{Strasberg2013Jan,Freitas2021Mar} and nonthermal-demon-like~\cite {Sanchez2019Nov,Whitney2016Jan}.

\begin{figure}[t]
    \includegraphics[width=\linewidth]{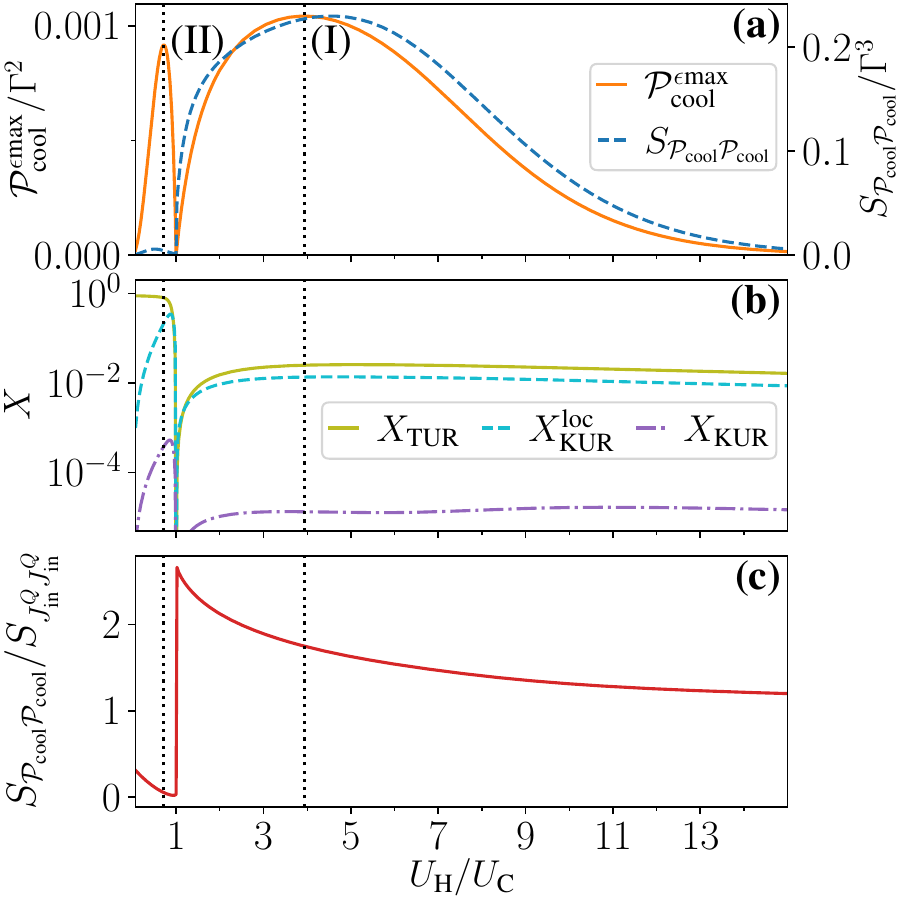}
    \caption{\label{fig:max power}
        (a) Cooling power maximized over $\eC$ and $\eW$, $\Pcool^{\epsilon \text{max}}$ (solid orange line), and the corresponding fluctuations in the cooling power, $\Scool$, (dashed blue line) as functions of $\UH$ (reproducing the results from Ref.~\cite[Fig.~3]{InfoTrajPaper}).
        (b) TUR and KUR performance quantifiers, as defined in Eqs.~\eqref{Q_TUR}, \eqref{Q_KUR},
        and \eqref{Q_KUR_loc}, when optimizing for cooling power.
        (c) Ratio of the cooling-power noise and input-current noise when optimizing for cooling power.
        The vertical dotted black lines indicate two specific parameter sets which will be called scenarios (I) and (II), see also Table~\ref{tab:params}. Parameters (in units of $\GC =\GH =\Gamma$): $\THot = 16$, $\TCold = 4$, $\T = 8$, $\dT = 2$, $\UC=12$, $\GR^{00} = \GL^{01} = \GL^{10} = 0$ and $\GL^{00} = \GR^{01} = \GR^{10} = 0.01$.
        These parameters are of the order of related experimental realizations~\cite{Thierschmann2015Oct}.}
\end{figure}

\subsection{Operation at optimized average cooling power}\label{sec_res_coolingpower}

\begin{table}[tb]

    \begin{tabular}{lcccc}
        \hline\hline
        Parameter  &  Scenario (I)  & Scenario (II) &  Regime (I)  & Regime (II)\\
        \hline
        $\TCold$  & 4  & 4  & 4  & 4 \\
        $\THot$   & 16 & 16 & 16 & 16\\
        $\bar{T}$ & 8  & 8  & 8  & 8 \\
        $\dT $    & 2  & 2  & variable  & variable \\
        $\mu_\C, \mu_\H, \mu_\L, \mu_\R$ & 0 & 0& 0 & 0\\

        $\GC,\GH$ & 1 & 1 & 1 & 1\\
        $\GR^{00}, \GL^{01}, \GL^{10}$ & 0 & 0 & 0 & 0\\
        $\GL^{00}, \GR^{01}, \GR^{10}$ & 0.01 & 0.01& 0.01 & 0.01\\
        $\UC$ & 12 & 12 & 12 & 12\\
        $\UH$ & 47.22 & 8.64 & 47.22 & 8.64\\
        $\eW$ & 7.59 & -9.88  & variable & variable \\
        $\eC$ & -7.08 & 1.24  & optimized & optimized \\
        $\eH$ & -4.96 & -14.38  & demon cond. & demon cond.\\
        \hline\hline
    \end{tabular}
    \caption{\label{tab:params}
        Parameters of the cases considered in this work, in units of $\Gamma$. In regimes (I) and (II), the cooling power $\Pcool$ is optimized over $\eC$ for each pair of values $(\dT, \eW)$ while $\eH$ is tuned to fulfill the demon condition.
    }
\end{table}

In Fig.~\ref{fig:max power}, we show the optimized cooling power, namely the optimal heat current flowing out of the right contact of the working substance, as a function of the ratio of the respective capacitive interaction strengths between dot W in the working substance and dots H and C in the resource region.
The optimization is done by tuning the level positions of dot C and W, for each ratio of interaction strengths $U_\mathrm{H}/U_\mathrm{C}$, while adjusting $\epsilon_\mathrm{H}$ such that $\Jin=0$.

Two parameter sets (see Table~\ref{tab:params} for the numerical values) are found in which the optimized cooling power reaches a local maximum as function of $U_\mathrm{H}/U_\mathrm{C}$, see Fig.~\subfigref{fig:max power}{a}.  One maximum is found for $\UH > \UC$ which we will refer to as \textit{scenario} (I) and another, slightly lower maximum, for $\UH < \UC$, that we will refer to as \textit{scenario} (II) in the following.
In Ref.~\cite{InfoTrajPaper}, these parameter regimes have been found to even have similar efficiencies, see also Supplemental Material~\cite{Suppl}. However, the two regimes were identified as being based on two rather different working principles: scenario (I) was found to exploit information and the system can be considered an autonomous Maxwell demon. In contrast, scenario (II) rather exploits the nonthermal properties of the resource distribution. Importantly, these two working principles have been shown to go along with very different \textit{fluctuation} characteristics. Indeed, tradeoff quantifiers involving the cooling power fluctuations shown in Fig.~\ref{fig:max power} are of different order of magnitude for the two scenarios. While scenario (I) has large cooling power fluctuations at the parameter set where the average cooling power has a maximum, the fluctuations in scenario (II) are strongly suppressed at the corresponding maximum in the cooling power.

\begin{figure}[tb]
    \includegraphics[width=\linewidth]{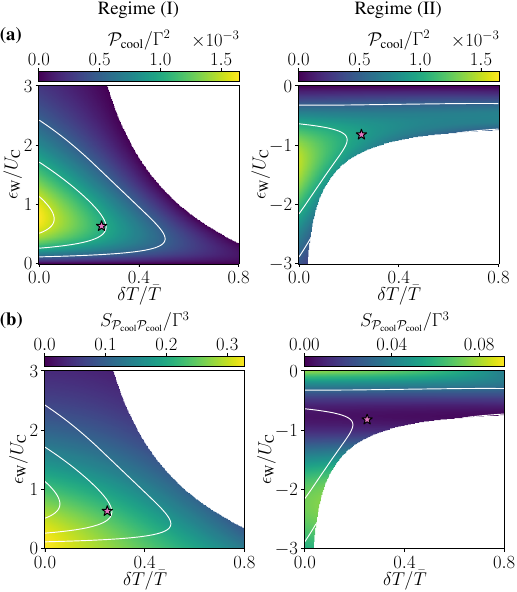}
    \caption{\label{fig:Pcool and Scool} (a) Average cooling power $\Pcool$ and (b) cooling power noise $\Scool$ as functions of $\eW$ and $\dT = \TL - \TR$. We choose the value of $\eC$ maximizing $\Pcool$ for each $\eW$ and $\dT$, while $\eH$ is always taken such that $\Jin = 0$. The parameters have the values of scenarios (I) and (II) (as given in Table \ref{tab:params}). The points corresponding to the exact parameters of scenarios (I) and (II) from Fig.~\ref{fig:max power} are indicated by a purple star. The solid white lines always indicate isolines of $\Pcool$.}
\end{figure}

We furthermore explore the behavior and performance of the system around those points by varying the temperature difference between the working substance reservoir, $\dT = \TL - \TR$, and the working substance dot energy, $\eW$. We will refer to these two cases as \textit{regimes} (I) and (II), since they have $\UH = \UH^\I$ and $\UH = \UH^\II$ respectively. However, in these two regimes, the value of $\eC$ is chosen to optimize the cooling power and $\eH$ is tuned to fulfill the demon condition $\Jin = 0$, see the last two columns of Table~\ref{tab:params}. Note that the range of values of $\dT$ we have chosen is such that $\TCold < \TR$, namely, the coldest temperature is in the resource region. Importantly, this is \textit{not} a necessary condition to achieve refrigeration under demon conditions. We show in the Supplemental Material \cite{Suppl} that, in regime (II), the reservoir being cooled down, namely reservoir R, can be the coldest---even colder than the cold resource reservoir.

In Fig.~\ref{fig:Pcool and Scool}, we plot the cooling power $\Pcool$ and its fluctuations $S_{\Pcool\Pcool}$, as functions of $\eW$ and $\dT = \TL - \TR$ in regimes (I) and (II).
White regions are always the ones where the demon condition cannot be fulfilled.
The points corresponding to the exact parameters of scenarios (I) and (II), as indicated by the dotted black lines in Fig.~\ref{fig:max power}---that is $\dT/\Gamma = 1$, with $\GC =\GH =\Gamma$, and $\eW = \eW^\I, \eW^\II$ as given in Table~\ref{tab:params}--- are marked by a purple star in the four panels of Fig.~\ref{fig:Pcool and Scool} and in subsequent figures. Solid white lines indicate isolines of $\Pcool$ in all density plots. Our choice of different ranges for values of $\eW$ in the left column ($\eW/\UC\in[0,3]$) and in the right column ($\eW/\UC\in[-3,0]$) is motivated by the sign of the cooling power. We are only interested in parameter ranges where the device operates as a refrigerator, namely when $\Pcool>0$. We therefore do not focus on situations where $\Pcool$ becomes negative, namely when $\eW$ becomes negative in regime (I) and when $\eW$ becomes positive in regime (II).
We see that the cooling power maxima are located at different level positions, leading to the differences in the operation identified for scenarios (I) and (II) in Ref.~\cite{InfoTrajPaper}. Furthermore, as expected, the cooling power is largest for $\delta T/\bar{T}\to 0$.

The fluctuations of the output cooling power are shown in Fig.~\subfigref{fig:Pcool and Scool}{b}. Interestingly, the maximum of the cooling power of regime (I) is located close to a maximum in the cooling power fluctuations as function of $\epsilon_\mathrm{W}$. Indeed, the yellow region of large cooling power in the left panel of Fig.~\subfigref{fig:Pcool and Scool}{a} largely overlaps with the yellow region of large cooling-power fluctuations in Fig.~\subfigref{fig:Pcool and Scool}{b}. The opposite is true for the maximum cooling power of regime (II). Here, the fluctuations have a minimum as function of  $\epsilon_\mathrm{W}$. Regions of large cooling-power fluctuations, indicated by yellow regions in the right panel of Fig.~\subfigref{fig:Pcool and Scool}{b} overlap with regions of \textit{small} cooling power in the right panel of Fig.~\subfigref{fig:Pcool and Scool}{a}, shown in dark blue.\footnote{The heat current in the resource $\Jtrans$ and the information currents in these two regimes, as well as fluctuations between all of those quantities are shown in the Supplemental Material~\cite{Suppl}.}

In the following, we discuss the implications that this has for the tradeoff quantifiers inspired by the thermodynamic and kinetic uncertainty relations in the two regimes of operation (I) and (II).

\subsection{Trade-off relations for engine precision}\label{sec_tradeoff}

The performance quantifiers $X_\TUR$, $X_\KUR$ and $X_\KUR^\loc$, as defined in Eqs.~\eqref{Q_TUR}, \eqref{Q_KUR}, and \eqref{Q_KUR_loc}, are plotted in Fig.~\subfigref{fig:max power}{b} for the parameters optimizing $\Pcool$. Both the TUR and KUR are much larger for scenario (II) than for scenario (I) and the TUR is even close to saturation in scenario (II). However, the local KUR is always much farther from saturation (reaching values up to $\simeq 0.35$) than the TUR (reaching values up to $\simeq 0.89$). See the Supplemental Material \cite{Suppl} for more details about the separate factors contributing to the performance quantifiers.

\begin{figure}[bt]
    \includegraphics[width=\linewidth]{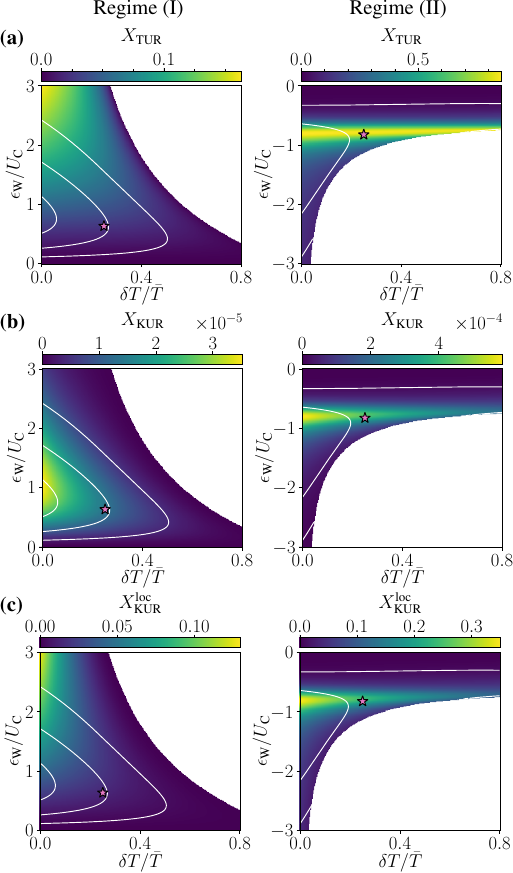}
    \caption{\label{fig:TUR and KUR}
        $X_\TUR$, $X_\KUR$ and $X_\KUR^\loc$ as functions of $\eW$ and $\dT = \TL - \TR$. The points corresponding to the exact parameters of scenarios (I) and (II), see Table~\ref{tab:params}, are indicated by a purple star. The solid white lines indicate isolines of $\Pcool$ in all the plots, see Fig.~\subfigref{fig:Pcool and Scool}{a}.}
\end{figure}

We now investigate these trade-off quantifiers more carefully in the parameter regimes around the ideal values of scenarios (I) and (II) using the density plots shown in Fig.~\ref{fig:TUR and KUR} for these two regimes. Here, we plot the performance quantifiers as functions of the temperature difference $\dT$ in the working substance and of the level position of dot W.
These plots confirm the first results of Fig.~\subfigref{fig:max power}{b} in that all factors $X$ have overall smaller values in regime (I) compared to regime (II) by about an order of magnitude.

We find that the optimal cases with respect to the cooling power lie close to maximum values of $X_\mathrm{TUR}$ and $X_\mathrm{KUR}^\mathrm{loc}$ in regime (II), while they lie in regions of low $X_\mathrm{TUR}$ and $X_\mathrm{KUR}^\mathrm{loc}$ in regime (I). This could, to some extent, be expected from the coincidence of high cooling power and low noise in scenario (II) and of high cooling power and high noise in scenario (I), even though care has to be taken since the $X$s are not determined by the fluctuations of $\Pcool$ alone.

This behavior is, however, slightly different for $X_\mathrm{KUR}$, which is again maximal for scenario (II), but is also in the vicinity of a maximum in scenario (I).  That this happens despite the fact that the cooling power fluctuations are not suppressed can be understood in the following way.
As seen from Fig.~\ref{fig:Pcool and Scool}, $\Pcool^2$ decreases with $\eW$ much faster from its maximum value than $\Scool$, and at the same time, the activity $\K$ does not vary much with $\eW$ in scenario (I), see Appendix~\ref{app:breaking loc KUR}.
The dependence of $Q_\KUR$ on $\eW$ is hence dominated by $\Pcool^2$ such that the maximum of $Q_\KUR$ approximately coincides with the maximum of $\Pcool$. Still, for scenario (I) the maximum value for $X_\mathrm{KUR}$ differs strongly from the value of $X_\mathrm{KUR}$ at maximum cooling power, as we discuss in the context of Fig.~\ref{fig:lassos} below.
Also, note that we find the KUR to be far from saturated in this setup for both regimes. Saturation of the KUR would require uniform tunneling rates~\cite{Macieszczak2024Jul}.
Here, however, the rates $\Gamma_{hwc} = -[W]^{hwc}_{hwc}$, where $W$ is the kernel of the master equation \eqref{eq:extmastereq} in the absence of counting fields ($\bm{\xi} = 0$), differ strongly.
Indeed, in all the configurations we consider, there is always at least one dot energy level that is negative, such that the empty state 000 decays fast to the state where that dot is occupied. This second state has, conversely, a lower decay rate because the occupation of the dot makes it harder to put an electron into another dot, to which it is capacitively coupled.

Interestingly, in regime (II), the TUR can be almost saturated (with $X_\TUR > 0.8$) by tuning $\eW$, even when increasing the temperature difference between the left and right reservoirs, see right panel in Fig.~\subfigref{fig:TUR and KUR}{a}. This, combined with the fact that the KUR remains far from being saturated, indicates that the device still operates in a ``close to equilibrium'' manner, despite the non-negligible temperature differences between the reservoirs.

It is furthermore insightful to investigate how the trade-off relations evolve with changing the level position $\eC$. Figure~\ref{fig:lassos} shows lasso plots for $X_\TUR$, $X_\KUR^\loc$, $X_\KUR$, and $\Scool$ versus $\Pcool$ for $\UH = \UH^\I$ and $\UH = \UH^\II$ respectively. Each point of the curves is obtained by maximizing $\Pcool$ over $\eW$ for a fixed value of $\eC$.\footnote{Note that this is not the same as in regimes (I) and (II), in which we maximize $\Pcool$ over $\eC$ while varying $\eW$.} By varying $\eC$, we cover the whole range of accessible cooling powers. We again confirm that all three trade-off factors are significantly better in the case where $\UH = \UH^\II$ compared to $\UH = \UH^\I$.

\begin{figure}[bt]
    \includegraphics[width=\linewidth]{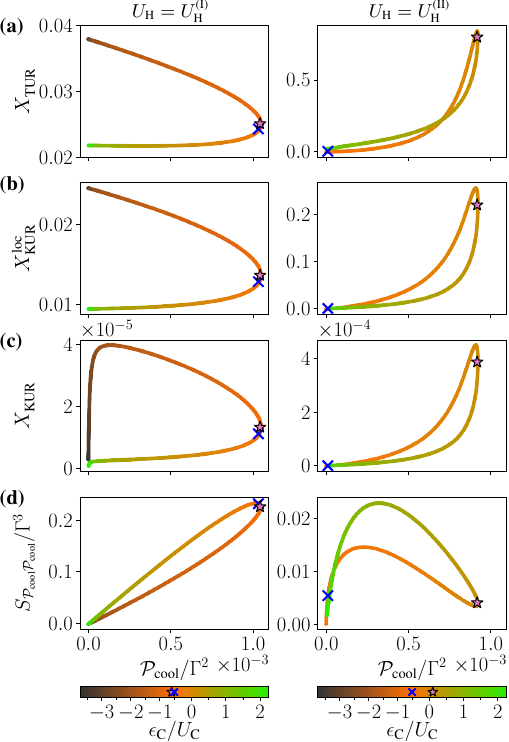}
    \caption{\label{fig:lassos}
        Lasso plots obtained by varying $\eC$ (as indicated by the colormap), maximizing $\Pcool$ over $\eW$ with $\eH$ chosen such that $\Jin=0$. Furthermore, $\UH$ is taken as in scenarios (I) and (II) respectively, and the other parameters are the same as in Fig.~\ref{fig:max power}.  The points corresponding to the exact parameters of scenarios (I) and (II) are indicated by a purple star. The blue crosses indicate the points at which $\eC = -\UC/2$.}
\end{figure}

We first analyze the case $\UH = \UH^\I$, shown in the left column of Fig.~\ref{fig:lassos}, for which the operation principle resembles the one of a Maxwell demon, meaning that information exchange between the working substance and the resource region was found to be crucial~\cite{InfoTrajPaper}.
We see that with increasing $\eC$ the cooling power first increases until the maximum cooling power (purple star) is achieved close to $\eC = -\UC/2$ (blue cross). Then the cooling power decreases again.
This is intuitively clear for a Maxwell demon's operating mode since the level position $\eC = -\UC/2$ of the dot attached to the cold contact allows for a good detection of the occupation of dot W.
Indeed, when $\eC$ significantly deviates from this value, the difference between $f_\C(\eC)$ and $f_\C(\eC+\UC)$ is suppressed, and the autonomous measurement and feedback mechanism associated with dot C and the cold resource reservoir on which the refrigeration relies stops working.
Also, the cooling power fluctuations, shown in the left panel of Fig.~\subfigref{fig:lassos}{d}, increase with increasing $\eC$ until a maximum value at $\eC\approx-\UC/2$ and then decrease again. Notably, they decrease faster when decreasing $\eC$ from close to $-\UC/2$ than when increasing it from $\eC\approx-\UC/2$. This is because in the former case dot C is occupied more often, which prevents dot H from being occupied, thus decreasing the fluctuations in the cooling power which would otherwise result from the detrimental contribution of the hot resource reservoir \cite{InfoTrajPaper}. Furthermore, $\Scool$ decreases faster with $\eC$ than $\Pcool$, see the left panel of Fig.~\subfigref{fig:lassos}{d}, which is at the origin of the increase in the three uncertainty relation quantifiers $X$ when decreasing $\eC$.

When choosing $\UH = \UH^\II$, the performances of the setup are very different, as can be seen in the right column of Fig.~\ref{fig:lassos}. This is in agreement with the previous conclusion that the operating principle is not mainly information-based. In other words, it does not solely rely on the cold resource reservoir, but also the energy exchange with the hot resource reservoir is beneficial in the regime of optimal cooling power.
We observe that not only the overall performance with respect to the trade-off factors is better in this case, but also the shape of the ``lassos'' is more intricate and they span a smaller range of $(X,\mathcal{P}_\mathrm{cool})$-pairs. This can be understood from the fact that $\Scool$ has a local minimum as function of $\eC$ when $\Pcool$ has a global maximum. The reason for this is that the exchange with both resource reservoirs is beneficial for the cooling operation around $\eC\approx 0$, thereby reducing the noise, see Sec.~\ref{sec_res_noise} below.
As a result of this local minimum in $\Scool$ the lasso plot of $\Scool$ is bent and the two branches of the lassos for the different factors $X$ are close to each other.

\subsection{Output versus input noise}\label{sec_res_noise}

\begin{figure}[tb]
    \includegraphics[width=\linewidth]{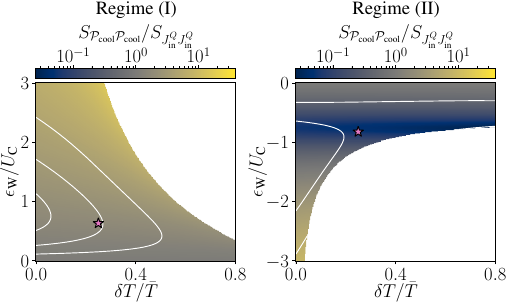}
    \caption{\label{fig:Sout/Sin}
        Ratio of the cooling power noise and input noise as a function of $\eW$ and $\dT = \TL - \TR$. The points corresponding to the exact parameters of scenarios (I) and (II) in Fig.~\ref{fig:max power}  (see also Ref.~\cite{InfoTrajPaper}) are indicated by a purple star. The solid white lines indicate isolines of $\Pcool$, see Fig.~\subfigref{fig:Pcool and Scool}{a}.}
\end{figure}

For the refrigerator considered here to work while operating under the demon condition, $\Jin=0$, fluctuations of the \textit{resource} are a requirement~\cite{Acciai2024Feb}. It is therefore important to understand to what extent these input fluctuations impact the precision of the output cooling power. In the linear-response regime of a thermal two-terminal engine, it has been shown that input fluctuations set a lower bound on the output fluctuations~\cite{Saryal2022Feb}. This is different in the more complex multi-terminal setup with capacitively coupled subsystems studied here. We show how the output cooling power fluctuations, $\Scool$, compare with the noise in the input heat current, $\Sin$, in Fig.~\ref{fig:Sout/Sin} for parameters in regimes (I) and (II).

The ratio between input and output noise is relatively uniform as a function of $\dT/\bar{T}$ and $\eW/\UC$ in regime (I). Input and output noise are always roughly of the same order of magnitude, the output noise being slightly larger than the input noise, in the same spirit as what was found for the two-terminal case~\cite{Saryal2022Feb}.

Also in the density plot of regime (II), input and output fluctuations differ by at most an order of magnitude, and the ratio has almost no dependence on the temperature difference. However, the overall behavior is different and we observe bigger differences in the ratio as function of $\eW$. In particular, the case of optimal cooling power is found in a regime, where $\Scool/\Sin$ has a minimum. At this minimum value, the output fluctuations are \textit{reduced} with respect to the input fluctuations.
This is an important insight for the operation of the device. While fluctuations in the input are required in order to make the refrigerator work (more details in the following Sec.~\ref{sec_res_cross}), this does not necessarily have a detrimental impact on the precision of the output. Note that this feature even persists when both resource reservoirs are hotter than reservoir R, which is to be cooled down, see Supplemental Material~\cite{Suppl}.

To understand these behaviors, the stochastic cycle analysis conducted in Ref.~\cite{InfoTrajPaper} is insightful, see also Ref.~\cite{Mayrhofer2021Feb} for a related configuration. We therefore briefly summarize it here and refer to Appendix~\ref{app:stoc cyc} for more details. The situation under study is described by the master equation \eqref{eq:extmastereq} (setting the counting fields to zero), which corresponds to a Markovian stochastic process between the occupation states $hwc$ of the dots. We now analyze trajectories in occupation-state space arising from this stochastic process.
Without loss of generality, we choose the empty state $000$ as a starting point and look at every time the system goes back to this state, calling the sequence of states of the three dots a stochastic cycle $\cC$. For any stochastic cycle, $\bcC$ denotes its time-reversed counterpart, namely the cycle where the jump sequence among dot states $hwc$ happens in reversed order. All relevant thermodynamic quantities, in particular the heat $Q_\alpha$ from reservoir $\alpha =$ C, H, L, R, can be defined at the cycle level \cite{InfoTrajPaper}, and will be denoted by $Q_\alpha(\mathcal{C})$ in the following. Moreover, note that $Q_\alpha(\bcC)=-Q_\alpha(\cC)$.

There are three kinds of thermodynamically relevant cycles~\cite{InfoTrajPaper}: $\CC$ involving only the cold resource reservoir, $\CH$ involving only the hot resource reservoir, and $\CHC$ involving both of them (see Appendix \ref{app:stoc cyc:def} and Ref.~\cite{InfoTrajPaper} for a precise definition). Together with their time-reversed counterparts, they form an ensemble of cycles which can be used to compute averages and fluctuations. In particular, the average current $J^Q_\alpha$ and noise $S_{J^Q_\alpha J^Q_\alpha}$, obtained from full counting statistics [Eqs.~\eqref{J_FCS} and \eqref{S_FCS}], can also be expressed based on the stochastic cycle quantities as \cite{Fiusa2025May, Fiusa2025Jun} (see Appendix \ref{app:stoc cyc:approx noise} for the justification of the approximation)
\begin{align}\label{eq:FCS-traj}
   J^Q_\alpha &= \frac{\mean{{Q}_{\alpha}}_\cC}{\tcyc}\ , &
   S_{J^Q_\alpha J^Q_\alpha} &\simeq \frac{\var_\cC({Q}_{\alpha})}{\tcyc}\ ,
\end{align}
where  $\mean{\bullet}_\cC$ and  $\var_\cC(\bullet)$ are, respectively, the average and variance over the stochastic cycles [see Eqs.~\eqref{eq:cycle_FCS}], and $\tcyc$ is the average duration of a cycle. As a consequence, $\Sin$ is given by the variance of $Q_\C(\cC) + Q_\H(\cC)$ and $\Scool$ is given by the variance of $Q_\R(\cC)$.

Let us start with discussing regime (I), where the device operates as a refrigerator for $\eW>0$. This means that the amount of heat $Q_\R(\CC)=\eW + U_{\C}$ exchanged during a cycle involving the cold resource reservoir, as well as the amount of heat $Q_\R(\CH)=\eW + U_{\H}$ exchanged during a cycle involving the hot resource reservoir, are always positive and hence beneficial for the cooling process. Yet, what makes the cooling process noisy is that it is always only one of them that is more probable than its reversed cycle. The reversed cycles of $\CH$ and $\CC$, namely $\bCH$ and $\bCC$ are however detrimental for cooling. As a result, the cooling power has large fluctuations as analyzed in Ref.~\cite{InfoTrajPaper}.

This is different in regime (II), where $\eW$ is negative so the sign of both $Q_\R(\CC) =\eW + \UC$ and $Q_\R(\CH) =\eW + \UH$ change with $\eW$.
This is important because, conversely, the entropy production $\Sigma(\cC)$ does not\footnote{For a given cycle, the entropy production satisfies $\Sigma(\cC)=\ln[\pi(\cC)/\pi(\bcC)]$, where $\pi(\cC)$ is the probability of occurrence of cycle $\cC$ and similarly for $\pi(\bcC)$~\cite{VandenBroeck2015Jan}.}, meaning that whether $\cC$ is more or less probable than its reversed $\bcC$ is independent of $\eW$, see Appendix~\ref{app:Sout/Sin}. In particular, in the range $-\UC < \eW < -\UH$, $Q_\R(\CC)$ and $Q_\R(\CH)$ have the same sign as their respective entropy productions, such that all the cycles contributing to the cooling ($Q_\R(\cC) > 0$) are more probable than their detrimental counterparts $\bcC$, which reduces the variance of $Q_\R(\cC)$. Outside of this area, only one of the cooling cycles is less probable than its detrimental counterpart, which increases the variance of $Q_\R(\cC)$. This explains why $\Scool$ exhibits a local minimum close to $\eW/\UC = -1$, the location of the noise reduction stripe observed in Fig.~\ref{fig:Sout/Sin}.

\begin{figure*}[tb]
    \includegraphics[width=\linewidth]{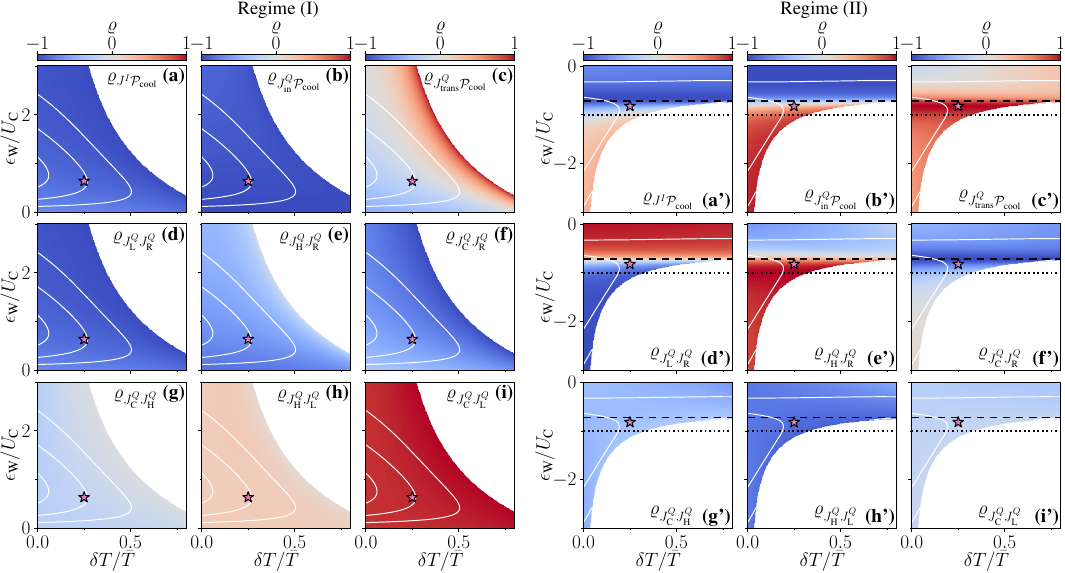}
    \caption{\label{fig:r}
        Correlation coefficient $\varrho$ from the full counting statistics [Eq.~\eqref{r}] for different pairs of currents as functions of $\eW$ and $\dT = \TL - \TR$: in regime (I) for the panels (a)-(i) and in regime (II) for panels (a')-(i'). The points corresponding to the exact parameters of scenarios (I) and (II), see Table~\ref{tab:params}, are indicated by a purple star. The solid white lines indicates isolines of $\Pcool$ in all the plots [see Fig.~\subfigref{fig:Pcool and Scool}{a}].  The horizontal dashed black line indicates $\eW = -\UH$ while the dotted black one indicates  $\eW = -\UC$.}
\end{figure*}

Finally, we see in Fig.~\subfigref{fig:max power}{c} that our findings about regimes (I) and (II) apply more generally to the broader parameter ranges $\UH > \UC$ and $\UH < \UC$, where only the latter exhibits a reduced noise in the cooling power compared to the noise in the input heat flow. This implies that multi-terminal devices fed by nonthermal resources do not only have the advantage to be able to perform useful thermodynamic tasks without absorbing heat on average; they can also have advantages with respect to precision, since there exist regimes where the output noise is suppressed with respect to the input noise in parameter regimes where the average output (here the cooling power) is large.

\subsection{Cross-correlations highlighting operation principles}\label{sec_res_cross}

Having studied the precision of the three-dot refrigerator, we now use the analysis of cross-correlations to further investigate its working principles.
Concretely, to analyze in more detail how the operation of the refrigerator differs in regimes (I) and (II), we study different Pearson correlation coefficients, see Eq.~\eqref{r}, given by cross-correlations between the key currents in the device.
In Fig.~\ref{fig:r}, we plot nine distinct Pearson correlation coefficients\footnote{The cross-correlations $S_{JJ'}$, entering the numerator of the Pearson coefficients, are shown in the Supplemental Material \cite{Suppl}.} as functions of $\eW$ and the temperature difference $\dT$: the correlation coefficients between the cooling power, $\Pcool = \JR$, and the input heat flow, the transverse heat flow and the information current respectively, as well as the six correlation coefficients between the heat flows of all possible pairs of reservoirs C, H, L, R.

We start by analyzing the working principles of the two scenarios (I) and (II) (and the parameter regimes in their vicinity) based on correlations between the cooling power and some resource currents of interest: the information current $J^I$, the input heat current $\Jin$ (which vanishes on average), and the transverse heat current in the resource region $\Jtrans$, see Fig.~\subfigref{fig:setup}{a}. As expected, the correlations between the information flow $J^I$ and the cooling power, shown in Figs.~\subfigref{fig:r}{a} and \subfigref{fig:r}{a'} differ significantly in the two regimes. In regime (I), which was identified as being the Maxwell-demon-like operation mode, $J^I$ and $\Pcool$ are strongly anticorrelated. Instead, in regime (II), which was identified as being the nonthermal-demon-like operation mode, the sign and magnitude of the correlations between $J^I$ and $\Pcool$ vary as a function of $\eW$; in particular, at the parameters of scenario (II) indicated by the purple star, the correlations are strongly suppressed.
We find a similar correlation pattern between $\Jin$ and the cooling power, $\Pcool$, in Fig.~\subfigref{fig:r}{b}, even though the average input heat flow is zero, $\Jin = 0$. There are always strong (anti-)correlations between $\Jin$ and $\Pcool$ in regime (I), while these correlations show large variations and a sign change in regime (II), with the maximum cooling power achieved in scenario (II) at a value of $\abs{\varrho_{\Jin\Pcool}}$ which is strongly suppressed, see Fig.~\subfigref{fig:r}{b'}.

Instead, in regime (II), the best refrigeration is obtained when there are large, positive correlations with the \textit{transverse} heat flow in the resource region, $\Jtrans$, as evidenced by Fig.~\subfigref{fig:r}{c'}. This is consistent with the previous conclusion that in scenario (II) the operating principle is nonthermal-demon-like and that a type of \textit{heat drag} is at play here~\cite{Whitney2016Jan,Bhandari2018Jul}. In regime (I), the correlations between $\Pcool$ and $\Jtrans$ are instead not significant, see Fig.~\subfigref{fig:r}{c}.

The study of the correlations can even be broken down to separate heat currents in the four contacts of working substance and resource and related to processes occurring at the trajectory level, thereby further highlighting the working principles of the two regimes. The results for this are shown in the lower two rows of Fig.~\ref{fig:r}; more details about the connection to the stochastic cycle analysis are provided in  Appendix~\ref{app:working_principle}.

Analyzing the cross-correlations between the left and right heat currents in the working substance, we see in Fig.~\subfigref{fig:r}{d} that $\JL$ and $\JR$ are strongly anticorrelated in regime (I), which is a signature of an information-based Maxwell-demon operating mode~\cite{Freitas2021Mar,Picatoste2024Aug}. In scenario (I), only one of the three dominating cycles (the three cycles with positive entropy production), namely $\CC$, cools down R. It functions by extracting heat from R and dumping it into L and C. In the two other dominating cycles, namely $\bCH$ and $\bCHC$, R receives heat from H and or L. This results in the strong anticorrelation between $\JL$ and $\JR$ [Fig.~\subfigref{fig:r}{d}]: whenever heat is transported into or out of L, the opposite is true for R. This, however, forces R to serve as a heat dump for H to fulfill the demon condition yielding a less precise cooling power than in scenario (II).

In regime (II) [Fig.~\subfigref{fig:r}{d'}], $\varrho_{\JL\JR}$ varies significantly and even changes sign as a function of $\eW$, meaning that the refrigeration in that case is not based on the creation of direct cross-correlations between $\JL$ and $\JR$. At the position of the purple star corresponding to scenario (II), $\varrho_{\JL\JR}$ is suppressed to $\varrho_{\JL\JR}\simeq -0.41$.
This happens despite having $\JL = -\JR$ for the average current due to the demon condition, and it confirms that the operating principle is hence nonthermal-demon-like~\cite {Whitney2016Jan,InfoTrajPaper} relying non-trivially on heat exchanges with the hot \textit{and} cold resource reservoirs.
From the stochastic cycle  analysis, one learns that in all three dominating cycles, heat is transferred from R into another reservoir by using additional energy supplied by a third reservoir. Whenever H is involved in one of these cycles, it serves as the energy supplier, and whenever C is involved, it receives heat. As a consequence $\JR$ is correlated with $\JH$ and anticorrelated with $\JC$, see Figs.~\subfigref{fig:r}{e'} and \subfigref{fig:r}{f'}. The situation for L is different, as it serves both as the energy supplier and the heat receiver in different cycles. This results in the vanishing correlations between $\JL$ and $\JR$ in scenario (II) as seen in Fig.~\subfigref{fig:r}{d'}. I.e., whenever heat is extracted from R, L could either receive heat, extract heat, or be bypassed. At the same time, all of the dominating cycles cool down R, allowing for the precise cooling power observed in Fig.~\ref{fig:Pcool and Scool}.

Finally, we aim to get further insights into the sign changes occurring in the correlation coefficients in regime (II). We therefore use the  generalization of Eq.~\eqref{eq:FCS-traj} to the cross-correlation between currents $J_\alpha^Q$ and $J_\gamma^Q$,
\begin{equation}
    S_{J_\alpha^Q J_\gamma^Q} \simeq \frac{\cov_\cC(Q_\alpha, Q_\gamma)}{\tcyc} = \frac{\mean{Q_\alpha Q_\gamma}_\cC}{\tcyc} - J^Q_\alpha J^Q_\gamma \tcyc\ ,
\end{equation}
see Appendix \ref{app:stoc cyc:approx noise} for the justification of the approximation. In the considered range of values of $\eW$ and $\dT$, the heat exchanges on the cycle level, $Q_\C(\cC)$, $Q_\H(\cC)$ and $Q_\L(\cC)$, do not change sign, and neither do the average heat currents $\JH,\JR > 0$ and $\JC, \JL < 0$. As a consequence, $\varrho_{\JC\JH}$, $\varrho_{\JH\JL}$,  and $\varrho_{\JC\JL}$ do not change sign in  Figs.~\subfigref{fig:r}{g'}, \subfigref{fig:r}{h'}, and \subfigref{fig:r}{i'}.

However,  $Q_\R(\CC) = \eW + \UC$ and $Q_\R(\CH) = \eW + \UH$ do change sign with $\eW$ in the parameter range of interest. Also, the probabilities of occurrence of the cycles---and therefore the respective weights of $Q_\R(\CC)$, $Q_\R(\CH)$ and $Q_\R(\CHC) = \UH - \UC < 0$ in the cycle average---depend on both $\eW$ and $\dT$.
Analyzing Figs.~\subfigref{fig:r}{b'}, \subfigref{fig:r}{d'}, and \subfigref{fig:r}{e'}, we find that $\varrho_{\Jin\Pcool}$, $\varrho_{\JL\JR}$, and $\varrho_{\JH\JR}$ always change sign at $\eW\simeq-\UH$ (indicated by a black dashed line), which is the value at which $Q_\R(\CH) = \eW + \UH$ changes sign. This means that the hot cycles $\CH, \bCH$ are the dominant contributions in the corresponding cross-correlations. By contrast, $\varrho_{J^I\Pcool}$ changes sign at $\eW \simeq - \UC$ [see black dotted line in Fig.~\subfigref{fig:r}{a'}], meaning that the dominant contribution to the cross-correlation comes instead from the cold cycles $\CC, \bCC$. This is consistent with the findings that only the refrigeration via the cold resource reservoir is powered by information in scenario (II) \cite{InfoTrajPaper}.
\footnote{Although less visible, $\varrho_{\JC\JR}$ and $\varrho_{\Jtrans\Pcool}$ also change sign.
These sign changes are the result of a more complex interplay between the sign changes of $Q_\R(\cC)$ and the variation of the cycle rates with $\eW$ and $\dT$.}

\section{Comparison with a noninteracting setup}\label{sec_scatt}

\begin{figure}[tb]
    \includegraphics[width=\linewidth]{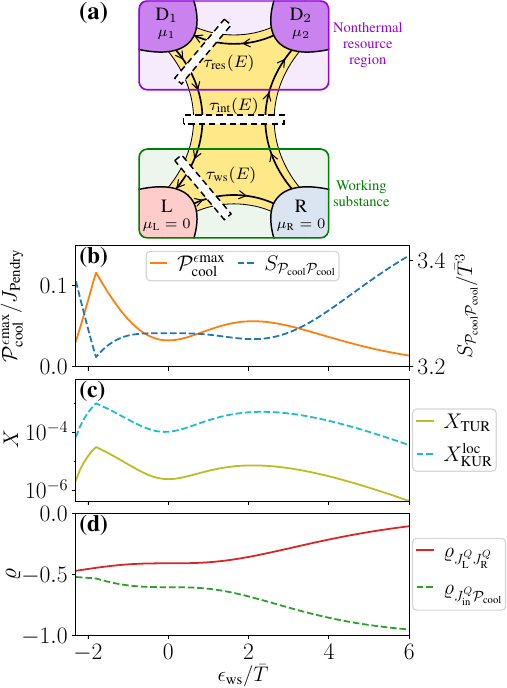}
    \caption{(a) Noninteracting setup: four electronic reservoirs $\D_1$, $\D_2$, L, R are connected by a scattering region with chiral quantum Hall channels. Scattering processes are encoded by transmission functions $\tau_i(E)$ that could be implemented by quantum point contacts (see main text).
        (b) Cooling power and noise in the cooling power as functions of $\epsilon_\text{ws}$.
        (c) Performance quantifiers $X_\TUR$ and $X_\KUR^\text{loc}$ as functions of $\epsilon_\text{ws}$.
        (d) Pearson correlation coefficients $\varrho_{\JL\JR}$ and $\varrho_{\Jin\Pcool}$ as functions of $\epsilon_\text{ws}$.  The temperatures are $T_{1}=T_{2}=1.2\bar{T}$, $T_{\L,\R}=\bar{T}\pm\delta T/2,\, \delta T=0.1\bar{T}$, while the onset $\epsilon_\mathrm{res}$ of the resource transmission function $\tau_\mathrm{res}(E)=\Theta(E-\epsilon_\mathrm{res})$ is optimized to maximize the cooling power.
    }
    \label{fig:scatt_setup_new}
\end{figure}

In this work, we have analyzed a quantum-dot based ``demonic" refrigerator exploiting a nonthermal resource (obtained here by mixing two thermal ones). A similar nonthermal demon has been proposed in a quantum Hall bar in Ref.~\cite{Sanchez2019Nov} and further characterized in Refs.~\cite{Hajiloo2020Oct,Acciai2024Feb}.
Here, we point out some crucial similarities and differences between the dot-based refrigerator of this work and the quantum-Hall-based refrigerator, see Fig.~\subfigref{fig:scatt_setup_new}{a}.

The quantum-Hall-based refrigerator is described in the framework of scattering theory, thereby neglecting Coulomb interactions beyond the mean-field level. Not relying on interactions, a first conceptual difference is hence that this device excludes the possibility of autonomous feedback, ruling out any information-based Maxwell-demon-like behavior.

The system of Fig.~\subfigref{fig:scatt_setup_new}{a} is modeled as follows: in the resource region, a transmission function $\tau_{\mathrm{res}}(E)$ mixes the two thermal distributions from reservoirs $\D_1$ and $\D_2$, which are given by Fermi functions.
In the working substance, an energy-dependent transmission function $\tau_\mathrm{ws}(E)$ is needed for a ``demonic'' operation, namely to have a negative local entropy production~\cite{Hajiloo2020Oct}. Finally, a possible interface transmission is present, which is, however, unimportant for the features we are interested in, and is therefore set to $\tau_\mathrm{int}(E)=1$.
The full scattering matrix describing the system is reported in Appendix~\ref{app:s_matrix}.

A further conceptual difference with the dot-based device is that the setup in Fig.~\subfigref{fig:scatt_setup_new}{a} allows particles to be exchanged between resource and working substance, hence the demon condition requires two equations: \emph{on average}, we require no energy \emph{and} particle exchange between the working substance and the resource, namely $J_\mathrm{in}^N=J_{\mathrm{in}}^E=0$. These conditions are fulfilled by fixing the chemical potentials $\mu_{1,2}$ of reservoirs $\D_{1,2}$, which are both measured with respect to the energy reference $\mu_\L=\mu_\R=0$.
Importantly, the setup of Fig.~\subfigref{fig:scatt_setup_new}{a} is such that the demon conditions are not affected by the choice of the features of the working substance, encoded in the transmission $\tau_{\mathrm{ws}}(E)$.
There are several possible choices for the transmission functions. As in the previous study~\cite{Acciai2024Feb}, we focus on sharp step functions $\tau_{i}(E)=\Theta(E-\epsilon_i)$, with $i=\text{res},\text{ws}$, since they are known to generate high output powers in similar thermoelectric devices~\cite{Whitney2014Apr,Whitney2015Mar}.  Such transmissions can be implemented by quantum point contacts.

In Figs.~\subfigref{fig:scatt_setup_new}{b}-\subfigref{fig:scatt_setup_new}{d}, we present the main quantities of interest, summarizing the behavior of the device, namely: the cooling power, its fluctuations, the KUR- and TUR-based figures of merit, and two Pearson correlation coefficients. Building on the results from Ref.~\cite{Acciai2024Feb}, we choose a parameter regime where the temperatures $T_{1},\, T_{2}$ in the resource are equal and larger than the base temperature $\bar{T}$ in the working substance, namely $T_{1}=T_{2}=1.2\bar{T}$, with $T_{\L,\R}=\bar{T}\pm\delta T/2$. The non-thermal character of the resource arises from a difference in electrochemical potentials $\mu_{1}\neq\mu_{2}$ between the two mixed resource contacts, the values of which are fixed by the \textit{two} demon conditions.

Regarding the onsets of the transmission functions $\epsilon_{\mathrm{ws}}$ and $\epsilon_{\mathrm{res}}$, we adopt a similar strategy as for the dot-based setup: for each value of $\epsilon_\mathrm{ws}$, we optimize $\epsilon_\mathrm{res}$ to achieve the maximum cooling power, $\mathcal{P}_\mathrm{cool}^{\epsilon\mathrm{max}}$. We express the latter as a fraction of Pendry's bound
$J_\mathrm{Pendry}=\pi\TR^2/12$,
representing the maximum heat current that can be extracted from the reservoir to be cooled down~\cite{Pendry1983Jul}.

We notice that, also in this system, the cooling power $\mathcal{P}_\mathrm{cool}^{\epsilon\mathrm{max}}$ has two maxima as a function of $\epsilon_\mathrm{ws}$, one of them occurring at the resonance condition $\epsilon_\mathrm{ws}=\epsilon_\mathrm{res}$. Which of the two maxima is the largest depends on the temperature difference $\delta T$~\cite{Acciai2024Feb}. Here, unlike in the dot-based demon, the fluctuations at the cooling power maxima are rather similar. Another significant difference is that the TUR and local KUR figures of merit are here much farther from saturation than in both scenarios identified in Fig.~\ref{fig:max power}. In the TUR case, this is due to the large entropy production in the resource region (driven by a significant chemical potential bias), which is necessary to achieve a large cooling power in the working substance. The low value of $X_{\mathrm{KUR}}^{\mathrm{loc}}$ can instead likely be attributed to the rather small temperature biases, as the KUR is typically closest to saturation far from equilibrium. Finally, the Pearson correlation coefficients are always negative, as required by free fermionic transport~\cite{Blanter2000Sep}. Unlike in the dot-based refrigerator, the heat currents $\JL$ and $\JR$ are never perfectly anticorrelated, and a minimum value of $\varrho_{\JL\JR}\approx -0.5$ is achieved.
Additionally, a large correlation between $\Jin$ and $\Pcool$ can be observed, but it remains far from perfect at the cooling power maxima. These seem to be the common features of demons exploiting nonthermal resources [see Figs.~\subfigref{fig:r}{b'} and~\subfigref{fig:r}{d'}], confirming
that the working principle is not that of an information-based Maxwell demon.
Understanding more precisely the working principles at play in all parameter regimes in this noninteracting setup would require further investigation, which is beyond the scope of this work.

\section{Conclusion}

We analyzed the fluctuations of a quantum-dot-based ``refrigerator", which is able to extract heat from a cold electronic reservoir in the ``working substance" of the device. The working principle of this refrigerator is special in the sense that it does not extract any energy on average from the resource region---hence seemingly operating like a demon. Performing a useful engine task (here cooling) while avoiding average energy transfer to the working substance can be of interest since it might reduce heating of the working substance.

The demonic action of this device can be identified as relying mostly on information exchange with the resource or on the nonthermal properties of the resource. Using full counting statistics combined with insights from stochastic trajectories, we characterized these two working principles based on the cross-correlations between different heat and information currents in the device. We furthermore showed that the information-based operation is typically noisier than the nonthermal operation. This is evident from their characterization in terms of tradeoff relations based on the thermodynamic and kinetic uncertainty relations, but also in how their cooling power fluctuations relate to the input fluctuations. Indeed, in the nonthermal operation mode, the output noise can be suppressed compared to the input noise.
We also pointed out differences and similarities of the studied quantum-dot refrigerator with respect to similar seemingly demonic devices realized in noninteracting coherent conductors.

By combining full counting statistics and trajectory analysis, our work provides insights into which processes of nanoscale engines are particularly affected by noise, and it indicates routes to improve the precision of nanoscale engines without requiring (large amounts of) average heat transfer. In the future, it could be interesting to study more generic types of nonthermal resources, as well as the impact of higher occupation numbers of the triple-dot system on the noise properties.

\acknowledgments
We thank Ludovico Tesser and Gabriel Landi for fruitful discussions. We acknowledge financial support from the Knut and Alice Wallenberg Foundation via the Fellowship program (J.M., M.A., J.S.), from the European Research Council (ERC) under the European Union’s Horizon Europe research and innovation program (101088169/NanoRecycle) (D.P., J.S.), from the PNRR MUR project No. PE0000023-NQSTI (M.A.),
the European Union's Horizon Europe research and innovation programme under the Marie Sklodowska-Curie grant agreement No. 101205255 FLUTE (M.A.), and from the Spanish Ministerio de Ciencia e Innovaci\'on via grants No. PID2022-142911NB-I00 and No. PID2024-157821NB-I00, and through the ``Mar\'{i}a de Maeztu'' Programme for Units of Excellence in R{\&}D CEX2023-001316-M (R.S.).

\section*{Data availability statement}

The data that support the findings of this article are openly available \cite{Data, Monsel2024Sep}.

\appendix

\section{Breaking the local KUR}\label{app:breaking loc KUR}

\begin{figure}[b]
    \includegraphics[width=\linewidth]{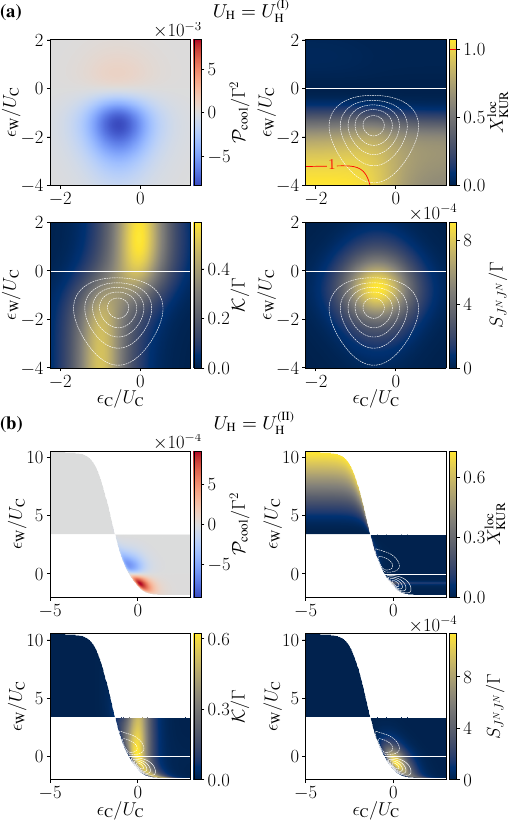}\\
    \caption{\label{fig:KUR loc}
    $X_\KUR^\text{loc}$ [Eq.~\eqref{Q_KUR_loc}], $\Pcool$, $\K$ and $S_{J^N J^N}$ in the $(\eC, \eW)$-plane with $\eH$ such that $\Jin=0$. All other parameters are chosen as in (a) scenario (I) and (b) scenario (II), see Table \ref{tab:params}. The white lines are isolines of $\Pcool$ and the solid red line highlights the area where $X_\KUR^\text{loc}> 1$ in panel (a).
}
\end{figure}
The TUR and KUR bounds are always satisfied in the dot-base refrigerator of this work, as it behaves according to a classical Markov process. This is not the case for the local KUR bound. Indeed,
the performance quantifier $X_\KUR^\loc$, defined in Eq.~\eqref{Q_KUR_loc}, is based on a local version of the KUR \cite{Palmqvist2025Oct} for noninteracting systems that is valid either close to equilibrium or in the tunneling regime. This bound can, however, be broken either due to interference effects in coherent quantum transport in the strong coupling regime, see Ref.~\cite{Palmqvist2025Oct}, or due to interactions between particles, which is outside of the validity range of the local KUR developed in Ref.~\cite{Palmqvist2025Oct}. Given the \textit{strong capacitive} couplings between the dots in the system studied here, we do not expect the inequality  $X_\KUR^\loc \le 1$ to hold.

Figure~\ref{fig:TUR and KUR} shows that in regime (II) the maximum in $\Pcool$ almost coincides with the maxima in $X_\text{TUR}$ and $X_\text{KUR}$ while this is not the case for regime (I). Focusing on the local KUR, we instead find that its maxima occur when $\Pcool < 0$ in regime (I), namely when no refrigeration takes place, as shown in Fig.~\subfigref{fig:KUR loc}{a}. Furthermore, we point out the fact that in this case, the local KUR can be broken, as highlighted by the red line in Fig.~\subfigref{fig:KUR loc}{a}. This is, however, in a parameter range where the device is not performing a useful task since $\Pcool < 0$. No breaking of the local KUR has been observed in regime (II), see Fig.~\subfigref{fig:KUR loc}{b}. We also note that there is not a single set of parameter values that maximizes $X_\text{TUR}$ and $X_\text{KUR}$ as $X_\text{TUR}$ tends to become constant for high enough $\eC$ (compared to $\UC$) and $X_\text{KUR}$ tends to become constant for low enough $\eC$.

The breaking of the local KUR observed here is related to the presence of interactions that can suppress the fluctuations through correlations, such that $S_{J^NJ^N}\leq \K_\R$ where $\K_\R$ is the local activity of the right reservoir (namely the activity with respect to hopping events involving the right reservoir). This causes the \textit{local} KUR to break in two ways: first, activity in other parts of the network can increase precision in the cooling power without increasing the \textit{local} activity. This means that the \textit{total} activity would be needed to bound the cooling-power noise. Second, the local KUR of Eq.~\eqref{Q_KUR_loc} uses particle current noise to approximate the local activity, something which can only work if $S_{J^NJ^N}\geq \K_\mathrm{R}$. Again, interactions are responsible for suppressing noise such that one can obtain $S_{J^NJ^N} \leq \K_\mathrm{R}$. Indeed, as displayed in Fig.~\ref{fig:KUR loc}, the activity and particle-current noise display rather different behavior, and it is clear that the violation of the local KUR occurs in a region where the $S_{J^N J^N}$ is small while the total activity is large.

There are multiple potential sources in a transport setting, which can cause correlations that can both increase or decrease noise, e.g., (effective) interactions or interference effects. However, in a classical Markovian system if all particle transfers are independent of each other, the total number of transfers into a reservoir is Skellam distributed~\cite{Skellam1946May}. This, in turn, means that the noise of a particle current into/out of a reservoir is equal to its local activity, which implies that a local KUR will be a valid bound. For the system studied here, it is the Coulomb interaction which causes the deviation from Skellam statistics; this is not necessarily the case in coherent quantum transport, where correlations between transfers can instead originate from, e.g., the Pauli blocking, even in the absence of interactions. A more detailed analysis of the impact of interactions on the local KUR in general is beyond the scope of this paper and will be presented elsewhere.\\

\section{Details about the theoretical model}\label{app:model}

\subsection{Full kernel for the full counting statistics}\label{app:FCS kernel}
Here, we give the full expression for the different terms of the extended kernel $W(\bm{\xi})$ of the generalized master equation \eqref{eq:extmastereq} used to compute the full counting statistics:
\begin{widetext}\small\allowdisplaybreaks
\begin{align}
    W_\C(\bm{\xi}) &=
    \begin{bmatrix}\
       -\GCop & \GCom \e^{i[\xi^A- \varepsilon_\C^0 \xi^E_\C- I_\C^0\xi^I]} & 0 & 0 & 0 & 0 \\
       \GCop \e^{i  [\xi^A + \varepsilon_\C^0 \xi^E_\C + I_\C^0\xi^I]} & -\GCom & 0 & 0 & 0 & 0 \\
       0 & 0 & -\GCip & \GCim \e^{i[\xi^A -  \varepsilon_\C^1 \xi^E_\C- I_\C^1\xi^I ]} & 0 & 0 \\
       0 & 0 & \GCip \e^{i [\xi^A + \varepsilon_\C^1 \xi^E_\C+ I_\C^1\xi^I]} & -\GCim & 0 & 0 \\
       0 & 0 & 0 & 0 & 0 & 0 \\
       0 & 0 & 0 & 0 & 0 & 0
    \end{bmatrix}\ ,
    \\\nonumber
    W_\H(\bm{\xi}) &=
    \begin{bmatrix}
        -\GHop & 0 & 0 & 0 & \GHom \e^{i[\xi^A-  \varepsilon_\H^0 \xi^E_\H- I_\H^0\xi^I]} & 0 \\
        0 & 0 & 0 & 0 & 0 & 0 \\
        0 & 0 & -\GHip & 0 & 0 & \GHim \e^{i[\xi^A- \varepsilon_\H^1 \xi^E_\H -iI_\H^1\xi^I]} \\
        0 & 0 & 0 & 0 & 0 & 0 \\
        \GHop \e^{i [\xi^A+ \varepsilon_\H^0 \xi^E_\H+ I_\H^0\xi^I]} & 0 & 0 & 0 & -\GHom & 0 \\
        0 & 0 & \GHip \e^{i[ \xi^A+ \varepsilon_\H^1 \xi^E_\H + I_\H^1\xi^I]} & 0 & 0 & -\GHim
    \end{bmatrix}\ ,
    \\\nonumber
    W_\L(\bm{\xi}) &=
    \begin{bmatrix}
        -\GLoop & 0 & \!\!\!\GLoom \e^{i [\xi^A -\varepsilon_\W^{00} \xi^E_\L]}\!\!\! & 0 & 0 & 0 \\
        0 & -\GLoip & 0 & \!\!\!\GLoim \e^{i[\xi^A-  \varepsilon_\W^{01} \xi^E_\L]}\!\!\! & 0 & 0 \\
        \!\GLoop \e^{i[ \xi^A + \varepsilon_\W^{00} \xi^E_\L]}\!\!\! & 0 & -\GLoom & 0 & 0 & 0 \\
        0 & \!\!\!\GLoip \e^{i[ \xi^A+ \varepsilon_\W^{01} \xi^E_\L]}\!\!\! & 0 & -\GLoim & 0 & 0 \\
        0 & 0 & 0 & 0 & -\GLiop & \!\!\!\GLiom \e^{i[\xi^A-  \varepsilon_\W^{10} \xi^E_\L]}\! \\
        0 & 0 & 0 & 0 & \!\!\!\GLiop \e^{i[ \xi^A+ \varepsilon_\W^{10} \xi^E_\L]}\!\!\! & -\GLiom
    \end{bmatrix}\ ,
    \\\nonumber
    \!\!W_\R(\bm{\xi}) &=
    \begin{bmatrix}
        -\GRoop & 0 &\!\! \!\!\!\!\!\!\GRoom \e^{i[\xi^A-\xi^N-\varepsilon_\W^{00}\xi^E_\R]}\!\!\! & 0 & 0 & 0 \\
        0 & -\GRoip & 0 &\!\!\!\!\!\! \GRoim \e^{i[\xi^A-\xi^N-i\varepsilon_\W^{01}\xi^E_\R]}\!\!\! & 0 & 0 \\
       \! \GRoop \e^{i[ \xi^A+ \xi^N   +i\varepsilon_\W^{00} \xi^E_\R]}\!\!\!\!\!\! & 0 & -\GRoom & 0 & 0 & 0 \\
        0 & \!\!\!\GRoip \e^{i[ \xi^A+ \xi^N + \varepsilon_\W^{01} \xi^E_\R]}\!\!\! & 0 & -\GRoim & 0 & 0 \\
        0 & 0 & 0 & 0 & -\GRiop & \!\!\!\GRiom \e^{i[\xi^A-\xi^N-\varepsilon_\W^{01} \xi^E_\R]}\!\! \\
        0 & 0 & 0 & 0 & \!\!\!\!\!\!\!\!\GRiop \e^{i[\xi^A + \xi^N + \varepsilon_\W^{10}\xi^E_\R]}\!\!\! & -\GRiom
    \end{bmatrix}\ .
\end{align}
\end{widetext}
We have defined
\begin{align}
    \Gamma_{\C/\H}^{w\pm} &= \Gamma_{\C/\H} f_{\C/\H}(\pm\varepsilon_{\C/\H}^w)\ ,\\\nonumber
    \Gamma_{\L/\R}^{hc\pm} &= \Gamma_{\L/\R}^{hc} f_{\L/\R}(\pm\varepsilon_\W^{hc})\ ,
\end{align}
and
\begin{align}
    \varepsilon_{\C/\H}^w &= \epsilon_{\C/\H} + wU_{\C/\H}\ , & \varepsilon_\W^{hc} &= \eW + c\UC + h\UH\ ,\\\nonumber
    I^w_\C &= -\ln\left[\frac{\p(w|00)}{\p(w|01)}\right]\ , &
    I^w_\H &= -\ln\left[\frac{\p(w|00)}{\p(w|10)}\right]\ .
\end{align}
Here, $I^w_\C$ and $I^w_\H$ correspond to the information acquired by the resource region  during the transitions $0w0 \to 0w1$ and $0w0 \to 1w0$, respectively \cite{Horowitz2014Jul, Takaki2022Nov, InfoTrajPaper}, and $\p(w|hc) = \p_{hwc}/(\p_{h0c} + \p_{h1c})$ is the probability of having $w=0,1$ electrons in dot W knowing that dots H and C contain $h$ and $c$ electrons, respectively.

\subsection{Influence of the tunneling asymmetry}\label{app:tunneling asym}

In the main text and in Ref.~\cite{InfoTrajPaper}, we consider the ideal situation of maximal tunneling asymmetry, where $\GR^{00}$, $\GL^{01}$, and $\GL^{10}$ are fully suppressed. Here, we study the impact of a nonideal tunneling asymmetry on the performance of the refrigerator. For simplicity, we continue  assuming that $\GL^{00} = \GR^{01}=\GR^{10}$  and $\GR^{00} = \GL^{01}=\GL^{10}$ and define the tunneling asymmetry \cite{Whitney2016Jan}
\begin{equation}\label{Lbd}
    \Lambda = \frac{\GL^{00}\GR^{10} - \GL^{10}\GR^{00} }{(\GL^{00} + \GR^{00})(\GL^{10} + \GR^{10})} = \frac{\GL^{00}\GR^{01} - \GL^{01}\GR^{00} }{(\GL^{00} + \GR^{00})(\GL^{01} + \GR^{01})}\ ,
\end{equation}
such that the ideal case corresponds to $\Lambda = 1$.

\begin{figure}[htb]
    \includegraphics[width=\linewidth]{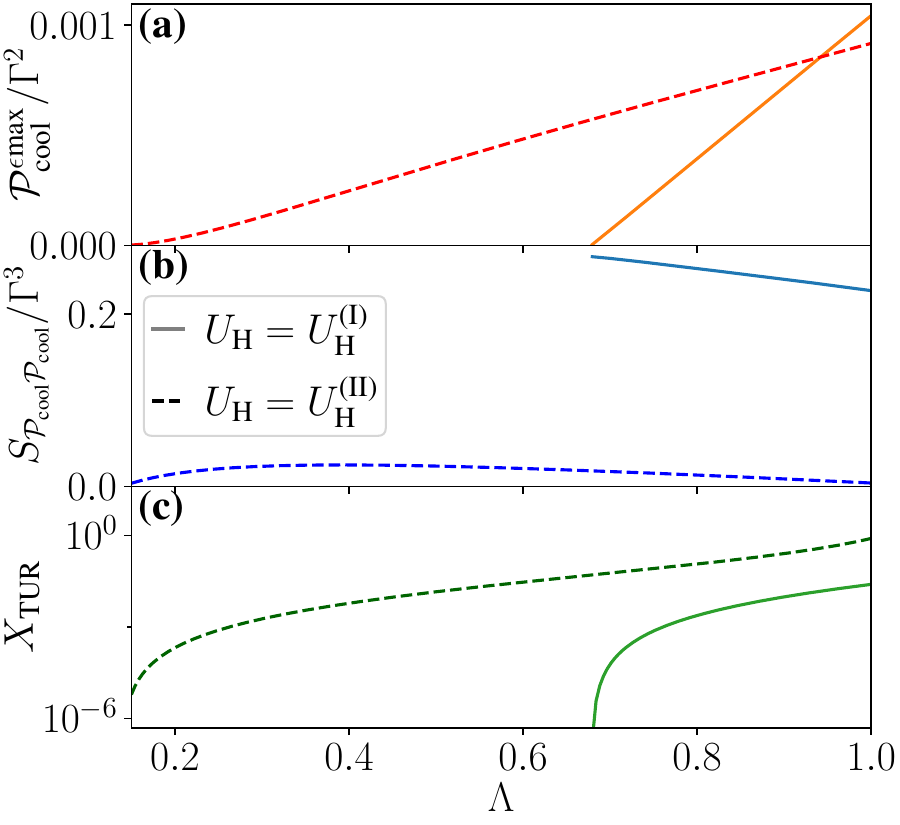}
    \caption{\label{fig:Lbd}
         Impact of the tunneling asymmetry. (a) Cooling power maximized over $\eC$ and $\eW$ while $\eH$ is chosen such that $\Jin = 0$, $\Pcool^{\epsilon\text{max}}$, (b)~the corresponding power fluctuations, $\Scool$, and (c)~performance quantifier $X_\TUR$ as functions of the tunneling asymmetry $\Lambda$ [Eq.~\eqref{Lbd}]. The solid lines were obtained with the same value of $\UH$ as in scenario (I) and the dashed lines with the value from scenario (II). To vary $\Lambda$, the value of the tunnel rates $\GL^{00} = \GR^{01}=\GR^{10}$ was kept fixed, while the value of $\GR^{00} = \GL^{01}=\GL^{10}$ was decreased down to zero. All the other parameters are the same as in Fig.~\ref{fig:max power}. Only the points with $\Pcool^{\epsilon\text{max}} > 0$ are plotted.
        }
\end{figure}

Figure~\ref{fig:Lbd} shows that, as expected, the performances degrade when the tunneling asymmetry decreases---the maximal cooling power $\Pcool^{\epsilon\text{max}}$ decreases, the fluctuations $\Scool$ increase, and $X_\TUR$ decreases---but refrigeration is nevertheless possible on a large range of nonideal values of $\Lambda$. We also see that $\Pcool^{\epsilon\text{max}}$ and $X_\TUR$ decreases much faster for $\UH > \UC$ [solid lines, same parameters as in scenario (I) in the limit $\Lambda \to 1$] than for $\UC > \UH$ [dashed lines, same parameters as in scenario (II) in the limit $\Lambda \to 1$]. Therefore, the latter case, in addition to its much higher precision, would also be more robust to experimental imperfections. In Fig.~\subfigref{fig:Lbd}{c}, we only plotted $X_\TUR$ but the other performance quantifiers, $X_\KUR$ and $X_\KUR^\text{loc}$, exhibit similar behaviors.

\section{Stochastic cycle analysis and its link to full counting statistics}\label{app:stoc cyc}

In this appendix, we briefly present the stochastic trajectory analysis of the setup as established in Ref.~\cite{InfoTrajPaper} and link it to the average currents and noises computed with the full counting statistics approach detailed in Sec.~\ref{sec:model:FCS} and Refs.~\cite{Fiusa2025May, Fiusa2025Jun}. We then use the framework of stochastic trajectories to gain further understanding of the fluctuations occurring in regimes (I) and (II).

\subsection{Definitions of cycle-related quantities}\label{app:stoc cyc:def}

\begin{figure}[t]
    \includegraphics[width=0.7\linewidth]{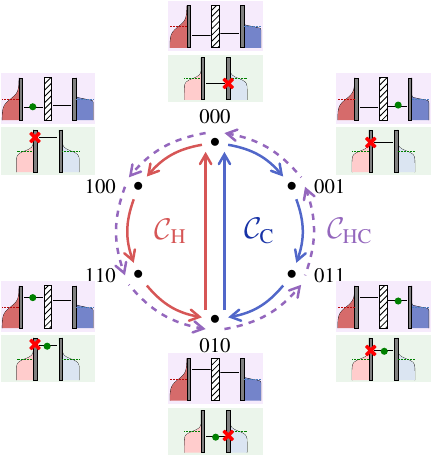}
    \caption{\label{fig:cycles}
        Sketch of the possible stochastic cycles, adapted from Ref.~\cite[Fig.~2]{InfoTrajPaper}.
    }
\end{figure}

The analysis of stochastic trajectories occurring in the state-space of the triple-dot system, as sketched in Fig.~\ref{fig:cycles}, can be reduced to an analysis of \textit{cycles} starting at state 000 and ending at state 000.
From a thermodynamic perspective, there are seven kinds of stochastic cycles (i.e., associated with distinct values of the thermodynamic quantities) \cite{InfoTrajPaper}: self-retracing sequences $\cC_0$ (resulting in no net exchanged heat, information, entropy, etc), cycles involving only the cold resource reservoir $\CC$ and $\bCC$, cycles involving only the hot resource reservoir $\CH$ and $\bCH$, and, finally, cycles involving both resource reservoirs $\CHC$ and $\bCHC$. We have denoted $\bcC$ the cycle corresponding to the reversed jump sequence compared to cycle $\cC$. In the following, we will denote $\mathcal{B} = \{\cC_0, \cC_\C, \cC_\H, \cC_{\H\C}, \bcC_\C, \bcC_\H, \bcC_{\H\C}\}$ this ensemble of cycles. The jump sequences (after removing any self-retracing part) of $\CC$, $\CH$, and $\CHC$ are given by
\begin{align}
    \gCC &= \GCop\GRoip\GCim\GLoom\ ,\nonumber\\
    \gCH &= \GHop\GRiop\GHim\GLoom\ ,\\\nonumber
    \gCHC &= \GHop\GRiop\GHim\GCip\GRoim\GCom\ ,
\end{align}
where the superscript on the left-hand side indicates the number of jumps. The probability $\pi(\cC)$ of cycle $\cC \in \{\CC, \bCC, \CH, \bCH, \CHC, \bCHC\}$ is given by
\begin{equation}\label{eq:raterelation}
    \pi(\cC)=  \rC \tcyc\ ,
\end{equation}
where $\tcyc$ is the average cycle time (including fluctuations)
\begin{equation}\label{t_cycle}
    \tcyc = \frac{1}{\p_{000}(\GCop + \GHop + \GLoop)}\ ,
\end{equation}
and $\rC$ the cycle rate, with
\begin{gather}
    \begin{aligned}
        \rCC &= \frac{\gH\gCC}{\g}\ ,& \rCH &= \frac{\gC\gCH}{\g}\ ,&  \rCHC &= \frac{\gCHC}{\g}\ ,\\
        \rbCC &= \frac{\gH\gbCC}{\g}\ ,& \rbCH &= \frac{\gC\gbCH}{\g}\ ,& \rbCHC &= \frac{\gbCHC}{\g}\ ,\label{cycle rates}
    \end{aligned}
\end{gather}
and
\begin{gather}
    \begin{aligned}
        \gC &= \GCom\GCim +\GCom\GRoim  +\GCim\GRoip\ , \\
        \gH &= \GHom\GHim +\GHom\GRiom  +\GHim\GRiop\ .
    \end{aligned}
\end{gather}
For details about the normalization factor $\gamma^5$, see Ref.~\cite{InfoTrajPaper}.
Any of the thermodynamic quantities of interest given in Sec.~\ref{sec:model:FCS} can be defined at the cycle level~\cite{InfoTrajPaper}, such that $\Xi(\cC)$ denotes the net amount of quantity $\Xi$ exchanged during cycle $\cC$, where $\Xi = N, A, I, E_\alpha, Q_\alpha$. For instance, during the cold cycle $\CC$, an amount of heat $Q_\R(\CC) = \eW + \UC$ is transferred from the right reservoir into dot W.

\subsection{Relation between full counting statistics and stochastic cycles}\label{app:stoc cyc:FCS link}

Let us consider a thermodynamic quantity of interest $\Xi = N, A, I, E_\alpha, Q_\alpha$, associated with the average current $J^{\Xi}$ and noise $S_{J^{\Xi} J^{\Xi}}$, which can be obtained from full counting statistics [Eqs.~\eqref{J_FCS} and \eqref{S_FCS}]. The average current and noise can also be expressed based on stochastic cycle quantities \cite{Fiusa2025May, Fiusa2025Jun}
\begin{subequations}\label{eq:cycle_FCS}
    \begin{align}
        \label{Jcyc} J^{\Xi} =\,& \frac{\mean{\Xi}_\cC}{\tcyc}\ , \\
        \label{eq:FCS_traj} S_{J^{\Xi} J^{\Xi}} =\,& \frac{\var_\cC(\Xi)}{\tcyc} + \frac{\Delta t_\cyc^2}{\tcyc}{(J^{\Xi})} ^2 \\\nonumber&- 2 \frac{\cov_\cC({\Xi}, {\tau_\cyc})}{\tcyc} J^{\Xi}\ .
    \end{align}
\end{subequations}
Here, we have defined the cycle average, variance and covariance
\begin{align}
    \label{eq:cycle_av_var_cov}
    \mean{\Xi}_\cC &= \sum_{\cC\in\mathcal{B}}\pi(\cC)\Xi(\cC)\ ,\nonumber\\
    \var_\cC(\Xi) &= \mean{\Xi^2}_\cC - \mean{\Xi}_\cC^2\ ,\\
    \cov_\cC(\Xi, \Phi) &= \mean{\Xi\, \Phi}_\cC - \mean{\Xi}_\cC\mean{\Phi}_\cC\ ,\nonumber
\end{align}
where $\Phi$ is a second stochastic quantity related to the cycle of interest. The average and variance of the duration $t_\cyc$ of a stochastic cycle are given by
$\tcyc$ and $\Delta t_\cyc^2$, and $\tau_\cyc$ is the duration of a cycle without counting the initial idle time spent by the system in 000. The cycle-time related quantities $\Delta t^2_\cyc$ and $\cov_\cC(\Xi, \tau_\cyc)$ can be computed analytically using full counting statistics, as explained in Ref.~\cite{Fiusa2025Jun}.  Note that $\tcyc = \mean{\tau_\cyc}_\cC +(\GCop + \GHop + \GLoop)^{-1}$ and $\Delta t_\cyc^2 = \var_\cC({\tau_\cyc}) + (\GCop + \GHop + \GLoop)^{-2}$, where the second terms correspond to the average and variance of the time the system spends in state 000.

Interestingly, we find that
Eq.~\eqref{eq:FCS_traj} even generalizes to covariances as
\begin{align}\label{Eq cross-corr}
    S_{J^{\Xi} J^{\Phi}} =\,&\frac{\cov_\cC(\Xi, \Phi)}{\tcyc} + \frac{\Delta t^2_\cyc}{\tcyc}J^{\Xi} J^{\Phi}\\\nonumber
    &- J^{\Phi} \frac{\cov_\cC(\Xi, \tau_\cyc)}{\tcyc}- J^{\Xi} \frac{\cov_\cC(\Phi, \tau_\cyc)}{\tcyc}\ .
\end{align}
This is a generalization of the results from Ref.~\cite{Fiusa2025Jun} and the proof and further analysis of this result, which are beyond the scope of this article, will be published elsewhere.

\begin{figure}[t]
    \includegraphics[width=\linewidth]{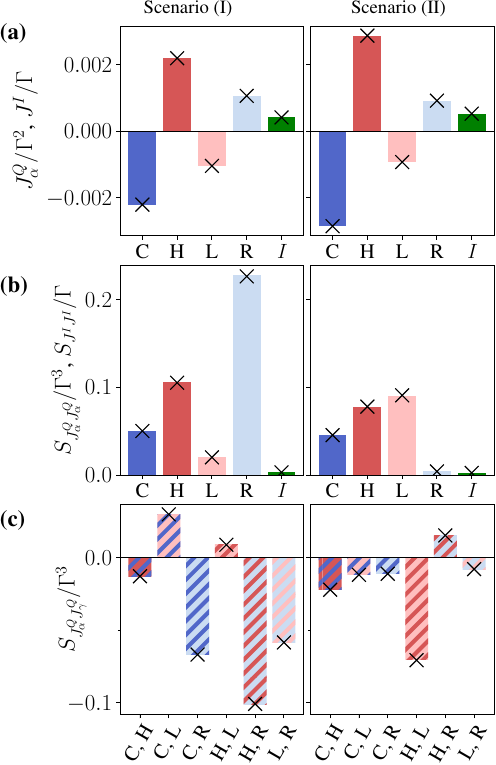}
    \caption{\label{fig:FCSvstraj}
        (a) Average currents, (b) noises, and (c) cross-correlations computed with full counting statistics (bars, see Sec.~\ref{sec:model:FCS}), and from the numerically generated trajectories (black crosses), using Eqs.~\eqref{eq:cycle_FCS} and \eqref{Eq cross-corr}, in scenarios (I) and (II). The labels on the $x$-axis indicate the reservoirs $\alpha$ and $\gamma$, except $I$ which indicates that the green bars are for the information current $J^I$. The color of the bars is a guide to the eye to help identify the corresponding reservoirs. The parameters are given in Table~\ref{tab:params}.
    }
\end{figure}

We show in Fig.~\ref{fig:FCSvstraj} that the average currents, current noises and cross-correlations estimated by averaging over numerically generated stochastic cycles using Eqs.~\eqref{Jcyc}, \eqref{eq:FCS_traj}, and \eqref{Eq cross-corr} are in excellent agreement with the analytical expressions of the same quantities, obtained from full counting statistics, as explained in Sec.~\ref{sec:model:FCS}.  The dataset, 65,000 trajectories for scenario (I) and 25,000 trajectories for scenario (II), is available at Ref.~\cite{Monsel2024Sep} (see also Appendix \ref{app:stoc cyc:traj link} discussing the connection between trajectories and cycles).

At the demon condition, $\Jin = 0$, the expressions of some of the heat current noises and cross-correlations simplify. In particular, from Eq.~\eqref{eq:FCS_traj}, we see that the noise in the input heat flow becomes
\begin{align}\label{Sin_cyc}
    \Sin &= \frac{\var_{\cC}(Q_\H + Q_\C)}{\tcyc} \\
    \nonumber&=  \sum_{{\cC=\CC, \CH, \CHC }}(Q_\H(\cC) + Q_\C(\cC))^2(\rC + r_{\bcC})\ ,
\end{align}
namely Eq.~\eqref{eq:FCS-traj} becomes exact.
Additionally, simplifying the analytical expression of $ S_{\Jin\Pcool}$ obtained from the full counting statistics, we find
\begin{align}\label{SJinJR}
    S_{\Jin\Pcool} =\,&  \frac{\cov_\cC(Q_\H + Q_\C, Q_\R)}{\tcyc} \\\nonumber&- \!\left[\UC\frac{\alpha_\H}{\gH}(\rCC - \rbCC)+\UH\frac{\alpha_\C}{\gC}(\rCH - \rbCH)\right]\!\Pcool\ ,
\end{align}
with
\begin{gather}
\begin{aligned}
    \alpha_\C &= {\GCom} + {\Gamma_\C^{1-}} + {\GRoim} + {\GRoip}\ ,\\
    \alpha_\H &= {\Gamma_\H^{0-}} + {\GHim} + {\GRiom} + {\GRiop}\ .
\end{aligned}
\end{gather}
Note that $\alpha_\C$ is the sum of the escape rates from 001 and 011 while $\alpha_\H$ is the sum of the escape rates from 100 and 110. The second term in Eq.~\eqref{SJinJR} corresponds to the cycle duration correlations in Eq.~\eqref{Eq cross-corr}, $- \Pcool {\cov_\cC(Q_\H + Q_\C, \tau_\cyc)}/{\tcyc}$.

\subsection{Approximation of the noise}\label{app:stoc cyc:approx noise}

While Eq.~\eqref{Jcyc} directly coincides with Eq.~\eqref{eq:FCS-traj}, Eq.~\eqref{eq:FCS_traj} shows that the current fluctuations are not directly equal to the variance of the corresponding stochastic thermodynamic quantity (first term in the equation). There are additional terms related to the variance of the duration of the cycle and the correlations between the cycle duration and the exchanged thermodynamic quantity (the second and third terms, respectively). However, we show in this Appendix that, in the parameter regimes considered in this work, the approximation for the noise considering the first term only, as done in Eq.~\eqref{eq:FCS-traj}, is well justified. Namely, the heat current noise is mostly due to the variance of the heat in the stochastic cycles.

\begin{figure}[t]
    \includegraphics[width=\linewidth]{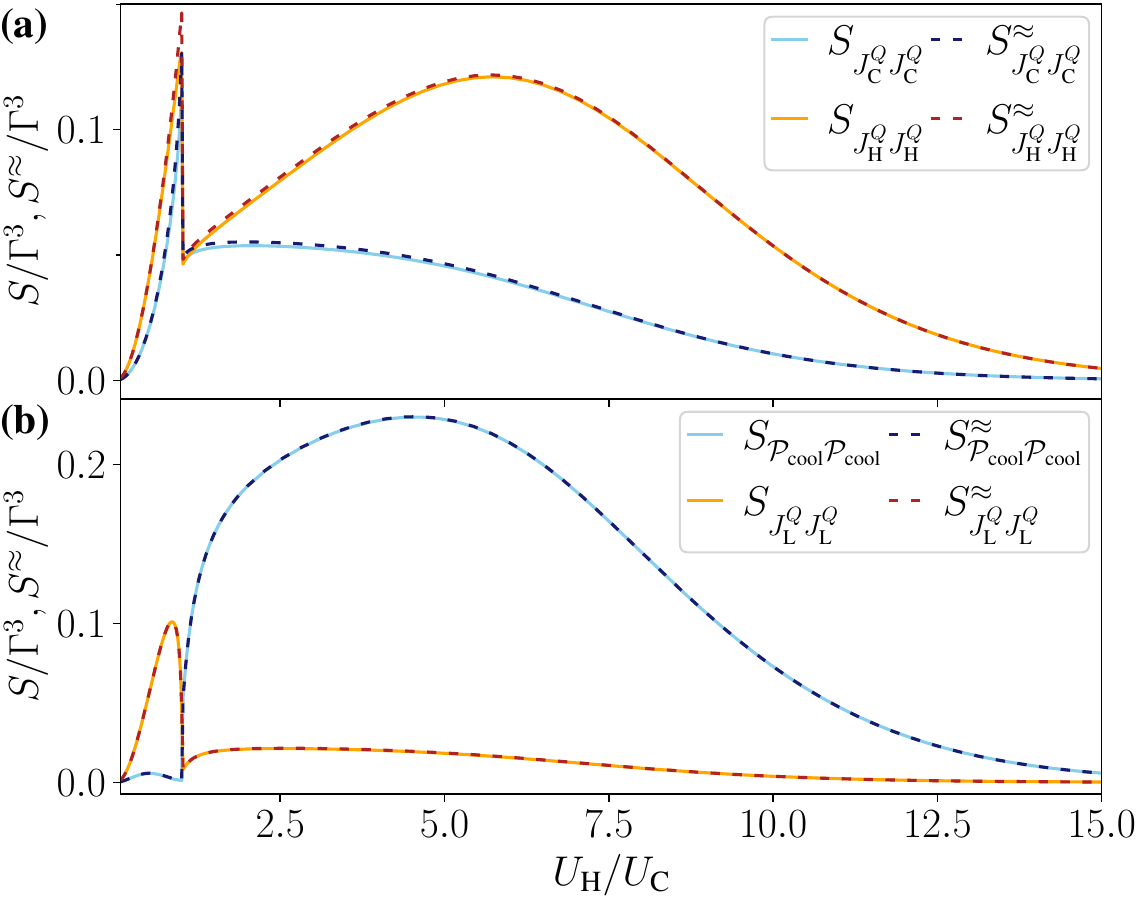}
    \caption{\label{fig:noise approx}
        Comparison between the full noise expression [Eq.~\eqref{S_FCS}] (solid lines) and the approximation from Eq.~\eqref{S_approx} (dashed lines) for (a) $S_{\JC \JC}$ and $S_{\JH\JH}$, and (b) $\Scool = S_{\JR\JR}$ and $S_{\JL\JL}$.
    }
\end{figure}

In Fig.~\ref{fig:noise approx}, we compare the full noise expression [Eq.~\eqref{S_FCS} or ~\eqref{Eq cross-corr}] to the approximation
\begin{equation}\label{S_approx}
    S^\approx_{J^{\Xi} J^{\Phi}} = \frac{\cov_\cC(\Xi, \Phi)}{\tcyc}\ ,
\end{equation}
which neglects correlations with the cycle durations in (see also the Supplemental Material \cite{Suppl} for more quantitative results in regimes (I) and (II)). We see that for the parameter regimes under consideration here, this approximation gives a good estimate of the noise and can therefore be used to interpret the behavior of the setup in regimes (I) and (II). We can however see a slight mismatch between $S$ and $S^\approx$, especially in Fig.~\subfigref{fig:noise approx}{a} close to $\UH= \UC$, due to the cycle time correlations, see Eq.~\eqref{Eq cross-corr}.

\subsection{Tight coupling between some of the currents}

Even when the demon condition is not fulfilled, we find tight couplings between some of the currents, meaning that for two quantities of interest $\Xi$ and $\Phi$, $J^{\Xi} = a J^{\Phi}$ and $S_{J^{\Xi} J^{\Xi}} = a^2 S_{J^{\Phi} J^{\Phi}}$ where $a$ can be expressed in terms of the system parameters (energies, temperatures and tunnel rates).
Using the expressions of the current and noise from Eqs.~\eqref{Jcyc} and \eqref{eq:FCS_traj}, we see that such a tight coupling exists if $\forall\,\cC\in \mathcal{B}$, $\Xi(\cC) = a\Phi(\cC)$. In that case, we get from Eq.~\eqref{Eq cross-corr} that any cross-correlation of $J^{\Xi}$ with another current $J'$ is given by $S_{J^{\Xi} J'} = a S_{J^{\Phi} J'}$.

A straightforward example is between the heat current from the left reservoir and the particle current. Indeed, the energy transport from dot W to the left reservoir happens at a single energy since $\GL^{01} = \GL^{10} = 0$, such that
\begin{equation}
    \JL = -\eW J^N\ , \quad S_{\JL\JL} = \eW^2 S_{J^NJ^N}\ .
\end{equation}

Another example is the entropy currents $J^s_\C$ and $J^s_\H$ with the respective heat currents $\JC$ and $\JH$. These entropy currents (defined in Refs.~\cite{Ptaszynski2019Apr, InfoTrajPaper}) correspond to the separate contributions from transitions in the cold and hot resource reservoirs to the information acquisition by the demon, with $J^I = -J^s_\C - J^s_\H$, and can be computed by replacing the counting field $\xi^I$ by $-\xi^s_\C$ in $W_\C(\bm{\xi})$ and $-\xi^s_\H$ in $W_\H(\bm{\xi})$, see Appendix~\ref{app:FCS kernel}.
Associated with the entropy currents $J_{\C,\H}^s$ is a cycle-related quantity $s_{\C,\H}(\cC)$ (again, defined in Refs.~\cite{Ptaszynski2019Apr, InfoTrajPaper}). Given that $s_{\C/\H}$ and $Q_{\C/\H}$ are related to transition with a single resource reservoir, they are proportional to each other, with $s_{\C/\H}(\cC) = \frac{s_{\C/\H}(\cC_{\C/\H})}{Q_{\C/\H}(\cC_{\C/\H})}Q_{\C/\H}(\cC)$, $\forall\,\cC\in \mathcal{B}$. Note however that since $\frac{s_{\C}(\cC_{\C})}{Q_{\C}(\CC)} \neq \frac{s_{\H}(\cC_{\H})}{Q_{\H}(\cC_{\H})}$ in general, $J^I= -J^s_\C - J^s_\H$ and $\Jin = \JC + \JH$  are not tightly coupled.

\subsection{Link between trajectories and cycles}\label{app:stoc cyc:traj link}

In Eq.~\eqref{eq:FCS-traj} and in Appendix \ref{app:stoc cyc:FCS link}, average values, variances, and covariances are taken with respect to stochastic cycles. Here, we make the link between these quantities and the stochastic trajectories $\gamma$ of fixed duration $t$, which is what is typically generated by numerical simulations \cite{Monsel2024Sep}. More concretely, since we are interested in the steady state, we want to express
$\lim_{t\to \infty} \frac{1}{t}\mean{\Xi(\gamma)}_{\gamma}$ and $\lim_{t\to \infty} \frac{1}{t}\mean{\Xi(\gamma)\Phi(\gamma)}_{\gamma}$ where $\mean{\bullet}_\gamma$ denotes the average over trajectories.

First, we can express the quantity of interest $\Xi$ along a trajectory as
\begin{equation}
    \Xi(\gamma) = \sum_{\cC\in\mathcal{B}} \Xi(\cC)M_\cC(\gamma)\ ,
\end{equation}
where $M_\cC(\gamma)$ is the number of occurrences of a cycle equivalent to $\cC$ when removing any self-retracing sequence along the trajectory $\gamma$. We are here neglecting contributions from the small initial and final portions of $\gamma$, which are not part of a closed cycle. This is justified since we are considering the long-time limit \cite{InfoTrajPaper}.
The occurrence of a cycle $\cC$ in the trajectory can be seen as a multinomial process. For long enough trajectories---such that the total number of cycles along any trajectory can be well approximated by $M_\cyc$, the closest integer to $t/\tcyc$---finding $M_\cC$ cycles in a trajectory is like the probability of having $M_\cC$ successes for $M_\cyc$ trials with a success probability $\pi(\cC)$ at each trial. Indeed, the process is Markovian so the next cycle is independent from the previous one. For each trial, we have in total seven mutually exclusive outcomes: $\cC_0, \cC_\C, \cC_\H, \cC_{\H\C}, \bcC_\C, \bcC_\H, \bcC_{\H\C}$. As a consequence,
\begin{align}
    \mean{M_\cC(\gamma)}_\gamma &= \pi(\cC) M_\cyc = \rC t,\\\nonumber
    \cov_\gamma(M_\cC, M_{\cC'}) &= \mean{M_\cC(\gamma)M_{\cC'}(\gamma)}_\gamma - \mean{M_\cC(\gamma)}_\gamma\mean{M_{\cC'}(\gamma)}_\gamma \\\nonumber
    &=  M_\cyc\pi(\cC)[\delta_{\cC\cC'}-\pi(\cC')]\\\nonumber
    &=  \rC t[\delta_{\cC\cC'} - r_{\cC'} \tcyc ]\ .
\end{align}
This gives the average current
\begin{equation}
    J^{\Xi} = \lim_{t\to \infty} \frac{1}{t}\mean{\Xi(\gamma)}_{\gamma} = \frac{\mean{\Xi}_\cC}{\tcyc}\ ,
\end{equation}
as in Eq.~\eqref{Jcyc}, but also the covariances
\begin{align}
    &\lim_{t\to\infty} \frac{\cov_\gamma(\Xi, \Phi)}{t} \nonumber\\\nonumber
    &\quad= \frac{1}{t}\left[\mean{\Xi(\gamma)\Phi(\gamma)}_\gamma - \mean{\Xi(\gamma)}_\gamma\mean{\Phi(\gamma)}_\gamma\right]\\\nonumber
    &\quad= \lim_{t\to \infty}\frac{1}{t}\sum_{\cC,\cC' \in \mathcal{B}}\Xi(\cC)\Phi(\cC')[ \mean{M_\cC(\gamma)M_{\cC'}(\gamma)}_{\gamma} \nonumber\\
    &\quad\hspace{3.75cm}-\mean{M_\cC(\gamma}\mean{M_{\cC'}(\gamma)}_{\gamma}] \nonumber\\
    &\quad= \sum_{\cC }\Xi(\cC)\Phi(\cC)\rC - \sum_{\cC, \cC'}\Xi(\cC)\Phi(\cC')\rC r_{\cC'} \tcyc \nonumber\\
    &\quad= \sum_{\cC}\Xi(\cC)\Phi(\cC)\rC
    - J^{\Xi} J^{\Phi} \tcyc\nonumber\\
    &\quad= \frac{\cov_\cC(\Xi, \Phi)}{\tcyc} \label{noise}\ .
\end{align}

\subsection{Output versus input noise dependence in $\eW$}\label{app:Sout/Sin}

\begin{figure}[t]
    \includegraphics[width=\linewidth]{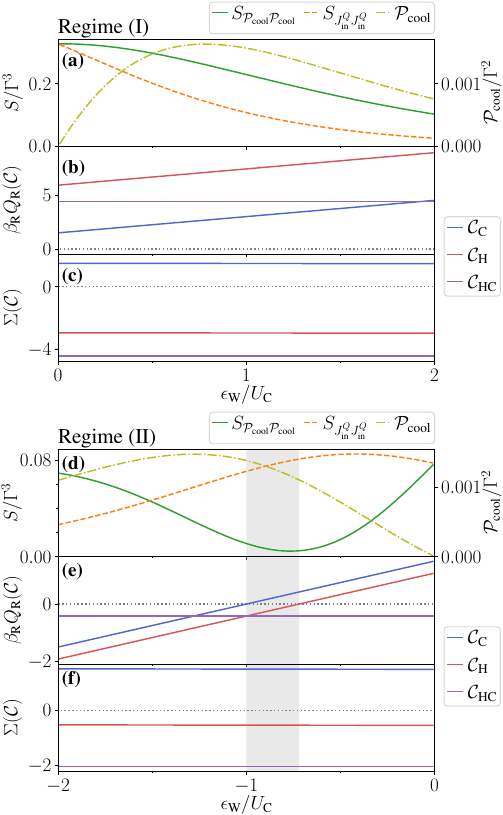}
    \caption{\label{fig:explain Scool/Sin II}
        (a), (d) Current noises $\Scool$ and $\Sin$ (left axis) and cooling power $\Pcool$ (right axis) as functions of $\eW$ in regimes (I) and (II), see Table~\ref{tab:params}, for $\dT/\Gamma = 0.05$.
        (b), (e) Corresponding heat $Q_\R$ extracted from the right reservoir (the reservoir in the working substance to be cooled) during cycles $\CC$, $\CH$ and $\CHC$.
        (c), (f) Corresponding entropy production $\Sigma$ during cycles $\CC$, $\CH$ and $\CHC$ (close to constant as function of $\eW$).
        The gray-shaded area in panels (d)-(f) indicates the range of values of $\eW$ where $Q_\R(\CC)$ is positive while $Q_\R(\CH)$ is negative.
    }
\end{figure}

Here, we provide more detailed insights concerning the ratio between input fluctuations and cooling-power fluctuations as discussed in Sec.~\ref{sec_res_noise}.
In Figs.~\subfigref{fig:explain Scool/Sin II}{a} and \subfigref{fig:explain Scool/Sin II}{d}, we show the numerator and denominator of the ratio $\Scool/\Sin$ as function of $\eW$ for $\dT/\Gamma = 0.05$, in regimes (I) and (II) respectively, that is along a vertical slice of each panel of Fig.~\ref{fig:Sout/Sin}. Figures~\subfigref{fig:explain Scool/Sin II}{b} and \subfigref{fig:explain Scool/Sin II}{e} display the heat $Q_\R(\cC)$ extracted from the right reservoir for each kind of cycle and Figs.~\subfigref{fig:explain Scool/Sin II}{c} and \subfigref{fig:explain Scool/Sin II}{f} the entropy production $\Sigma(\cC) = -\sum_\alpha Q_\alpha(\cC)/T_\alpha$ \cite{InfoTrajPaper}.

We notice that in both regimes $\Sin = \Scool$ at $\eW = 0$. This can also be proven analytically using Eqs.~\eqref{eq:cycle_FCS}. At the demon condition, $\Jin = 0$, we have $\Pcool = -\JL$ which vanishes at $\eW = 0$ since $Q_\L(\cC) = 0$ for all cycles at that point (see Table~\ref{tab:thermo}). As a result of the vanishing average currents, Eq.~\eqref{S_approx} is no longer an approximation, but the exact expression for $\Sin$ and $\Scool$, and at $\eW$, $[Q_\C(\cC) + Q_\H(\cC)]^2 = Q_\R(\cC)^2$ for all cycles (see Table~\ref{tab:thermo}), such that the ratio $\Scool/\Sin$ equals 1.

In regime (I), the refrigeration occurs for $\eW> 0$. Therefore, based on the expressions in Table~\ref{tab:thermo}, $[Q_\C(\cC) + Q_\H(\cC)]^2 < Q_\R(\cC)^2$ for all cycles, such that $\Sin < S^\approx_{\Pcool\Pcool}$ [Eqs.~\eqref{Sin_cyc} and \eqref{S_approx}]. Since we have seen in Appendix \ref{app:stoc cyc:approx noise} that $S^\approx_{\Pcool\Pcool}$ gives a very good approximation of the noise in the cooling power, we can state that $\Scool/\Sin \ge 1$ in regime (I), and therefore the output is always noisier than the input.

We now focus on regime (II). While $\Sin$ overall increases with $\eW$, $\Scool$ has a more complex behavior with a significant dip close to $\eW/\UC = -1$. This is exactly the parameter range of the noise-reduction ``stripe" observed in Fig.~\ref{fig:Sout/Sin}.
The suppressed ratio between output and input noise is hence related to a distinct suppression of the cooling-power fluctuations. In the following, we analyze the behavior of the trajectories contributing to this feature.

We have shaded in gray in Figs.~\subfigref{fig:explain Scool/Sin II}{d}-\subfigref{fig:explain Scool/Sin II}{f} the range of values of $\eW$, $-\UC < \eW < -\UH$ for which both $Q_\R(\CC)$ and $Q_\R(\CH)$ have the same sign as the related entropy production, respectively $\Sigma(\CC)$ and $\Sigma(\CH)$. In this range, the heat extracted from the colder right reservoir, $Q_\R(\CC)$, during cycle $\CC$ is positive (hence beneficial for the cooling process) and $\Sigma(\CC)$ is positive as well, making this cycle more probable than its reversed (non-beneficial) counterpart. At the same time, $Q_\R(\CH)$ is negative, but the occurrence of $\CH$ is suppressed since the related entropy production is also negative.  This leads to suppressed fluctuations of the cooling power.
Outside of this area, one of the cooling cycles is less probable than its detrimental counterpart, which increases the variance of $Q_\R(\cC)$. Note that while $Q_\R(\cC)$ can change sign when $\eW$ increases (for $\cC = \CC, \CH$), $\Sigma(\cC)$ does not. Indeed, $Q_\R(\CHC)$ and $\Sigma(\CHC)$ are independent of $\eW$ while $\Sigma(\CC)$ and $\Sigma(\CH)$ (see expressions in Table~\ref{tab:thermo}) are still close to constant. As $\TL \simeq \TR$, the term depending on $\eW$ is small compared to the interaction contribution, and therefore $\Sigma(\CC)$ is always positive and $\Sigma(\CH)$ is always negative, as shown in Fig.~\subfigref{fig:explain Scool/Sin II}{f}. Finally, comparing $\Sin$ and $\Scool$ , we get
\begin{align}
    \Scool - \Sin
    \simeq \,&\Scool^\approx - \Sin \nonumber\\
    \simeq\,& \eW(\eW + 2\UC)(\rCC + \rbCC)\\\nonumber
    & + \eW(\eW + 2\UH)(\rCH + \rbCH)\ .
\end{align}
Since in regime (II), the dominant process (in term of cycle rate) is the one involving only the hot resource reservoir \cite{InfoTrajPaper} (see also Supplemental Material \cite{Suppl}), the main contribution to $\Scool - \Sin$ is $\eW(\eW + 2\UH)(\rCH + \rbCH)$ which is negative for $-2\UH < \eW < 0$. In Fig.~\subfigref{fig:explain Scool/Sin II}{d}, the curves for $\Sin$ and $\Scool$ indeed cross in the vicinity of $-2\UH/\UC = -1.44$.

\begin{table}[t]
    \begin{tabular}{cccccc}
        \hline\hline
        \hspace{0.2cm}Cycle \hspace{0.2cm}\,& \hspace{0.2cm} $Q_\C$ \hspace{0.2cm}\,& \hspace{0.2cm}$Q_\H$\hspace{0.2cm}\, & \hspace{0.2cm}$Q_\L$\hspace{0.2cm}\, &\hspace{0.2cm} $Q_\R$\hspace{0.2cm}\,&$\Sigma$\,\\
        \hline
        $\CC$   & $-\UC$ & $0$    & $-\eW$ & $\eW {+} \UC$ & $\displaystyle\Delta\beta_{\C\R}\UC {-} \Delta\beta_{\R\L}\eW$ \\
        $\CH$   & $0$    & $-\UH$ & $-\eW$ & $\eW {+} \UH$ & $\displaystyle-\Delta\beta_{\R\H}\UH {-} \Delta\beta_{\R\L}\eW$ \\
        $\CHC$  & $\UC$  & $-\UH$ & $0$    & $\UH {-} \UC$ & $\displaystyle-\Delta\beta_{\C\R}\UC {-} \Delta\beta_{\R\H}\UH$ \\
        \hline\hline
    \end{tabular}
    \caption{\label{tab:thermo}
        Thermodynamic quantities at the cycle level \cite{InfoTrajPaper}. We have denoted $\Delta \beta_{\alpha\gamma} = 1/T_\alpha - 1/T_\gamma$, with $\alpha, \gamma =$ C, H, L, R. The values for the reversed cycles are given by $Q_\alpha(\bcC) = - Q_\alpha(\cC)$, $\Sigma(\bcC) = - \Sigma(\cC)$.}
\end{table}

\subsection{Pearson correlation coefficient}\label{app:Pearson}

Our definition of the Pearson correlation coefficient in Eq.~\eqref{r} is similar, but not exactly equal, to the Pearson correlation coefficient defined in Ref.~\cite{Freitas2021Mar}. Indeed, in Ref.~\cite{Freitas2021Mar}, the Pearson correlation coefficient is defined from stochastic quantities, namely as
\begin{equation}
    \mathcal{R}_{\JL\JR} = \frac{\cov_\gamma(Q_\L, Q_\R)}{\sqrt{\var_\gamma(Q_\L)\var_\gamma(Q_\R)}}
\end{equation}
for stochastic trajectories $\gamma$ of a given duration $t$.

We are interested in the steady state (and therefore  in trajectories with $t \to \infty$) such that, using the results from Appendix \ref{app:stoc cyc:traj link}, we  write
\begin{equation}
    \mathcal{R}_{J^{\Xi} J^{\Phi}} = \frac{\cov_\cC(\Xi, \Phi)}{\sqrt{\var_\cC(\Xi)\var_\cC(\Phi)}}\ ,
\end{equation}
for two currents $J^{\Xi}$ and $J^{\Phi}$. From Eqs.~\eqref{eq:FCS_traj} and \eqref{Eq cross-corr}, we see that $\varrho_{J^{\Xi} J^{\Phi}}$ and $\mathcal{R}_{J^{\Xi} J^{\Phi}}$ are not exactly equal. However, from the discussion in Appendix~\ref{app:stoc cyc:approx noise}, we expect them to be approximately equal for the parameter regimes we are interested in, and we show that this is indeed the case by plotting $\mathcal{R}_{\JL\JR}$, $\mathcal{R}_{\Jin\Pcool}$ and $\mathcal{R}_{\Jtrans\Pcool}$ in the Supplemental Material \cite{Suppl}. Note that the coefficient $\varrho$ defined in Eq.~\eqref{r} can be computed from current-noise measurements, while $\mathcal{R}$ would require measuring stochastic thermodynamic quantities along a very large number of trajectories. This means that $\varrho$ would be easier to obtain experimentally.

\section{Working principles of scenarios (I) and (II)}\label{app:working_principle}

\begin{figure}[t]
    \includegraphics[width=\linewidth]{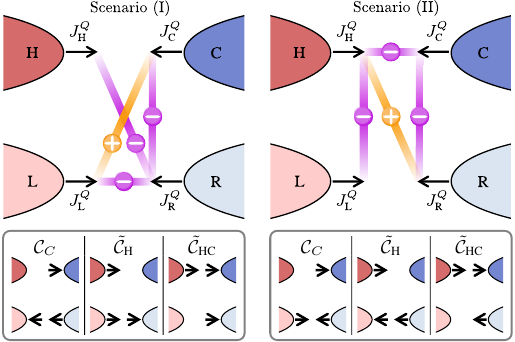}
    \caption{\label{fig:working_principle}
    Sketch of the four reservoirs H, C, L, and R forming the demon and working substance, respectively, along with the cross correlations between the heat currents out of the contacts, based on Fig. \ref{fig:r}. Heat currents out of the reservoirs are represented by the black arrows, and their cross-correlations are represented by the orange and purple beams connecting the different arrows corresponding to the pair of currents having a positive or negative Pearson coefficient. The insets show the direction of the nonzero heat transfer for the cycles with positive entropy productions (that is more probable than their reversed counterparts) in each scenario.
     }
\end{figure}
Here, we provide more details about the working principles behind scenarios (I) and (II) based on the correlations observed between the heat currents out of the different contacts, quantified by the Pearson coefficients presented in Fig. \ref{fig:r}. To do this, we display the positive and negative Pearson coefficients for the pairs of heat currents as orange and purple beams in the sketch of the device of Fig.~\ref{fig:working_principle}. To begin with, we note that a positive Pearson coefficient between two currents indicates that the two heat flows tend to have the same sign at a fluctuating level, whereas a negative coefficient implies the opposite. If the Pearson coefficient takes the extreme values  $\{1,-1\}$, the correlation or anticorrelation is perfect.

For scenario (I), Fig. \ref{fig:working_principle} indicates that the heat current out of reservoir R is anticorrelated with those out of H, C, and L. This behavior can be explained by studying the cycles with positive entropy production since these are the ones dominating the dynamics. In scenario (I),  $\mathcal{C}_\mathrm{C}$ is the only such cycle, which cools down R. It functions by letting heat flow from R into L and C. By contrast the cycles $\bcC_\mathrm{H}$ and $\bcC_\mathrm{HC}$ work by letting heat flow from H and L into R, or from H into C and R. As a consequence, the heat flows out of L and R usually have opposite signs, explaining why $\varrho_{\JL\JR}\approx-1$. Moreover, the anticorrelation between $\JH$ and $\JR$ stems from the fact that $\R$ serves as a heat dump for $\H$ to satisfy the demon condition. This is counterproductive to the goal of cooling down of R, resulting in the substantial noise in the cooling power of scenario (I).

We observe rather different correlations in scenario (II), and from Fig. \ref{fig:r}, we learn that $\JR$ is positively correlated with $\JH$ and anticorrelated with $\JC$. Here, the three dominating cycles can be viewed as refrigeration cycles where heat is extracted from R into another reservoir by supplying additional energy from a third reservoir. In particular, L serves as both the energy supplier in $\CC$ and as the reservoir receiving the heat in $\bCH$. This explains the vanishing correlations between $\JL$ and $\JR$ in scenario (II), whenever heat is extracted from R, heat could flow into L, out of L, or bypass L. On the other hand, since H only serves as the energy supplier, and C as the heat receiver whenever they are involved in one of the dominating cycles, they display stronger negative and positive correlations with $\JR$, respectively. This working principle allows L to serve as a heat dump, maintaining the demon conditions, and for all of the dominating cycles to cool down R in scenario (II). This is the reason for the more precise cooling power in scenario (II) compared to scenario (I).

\section{Scattering matrix of the noninteracting setup}\label{app:s_matrix}

For completeness, we report here the full scattering matrix $\mathscr{S}(E)$ that completely describes the transport properties of the noninteracting setup considered in Sec.~\ref{sec_scatt}, namely the currents $J_\alpha^\Xi$, and the corresponding fluctuations $S_{J_\alpha^\Xi J_\beta^\Xi}$, with $\alpha,\beta\in\{\L,\R,1,2\}$ [cf. Fig.~\subfigref{fig:scatt_setup_new}{a}], and $\Xi\in\{N,Q\}$. Up to irrelevant phase factors in the scattering matrix, we have 
\begin{widetext}
{\small\allowdisplaybreaks\setlength{\arraycolsep}{2pt}
\begin{gather}
\mathscr{S}(E)=
    \begin{pmatrix}
        -\sqrt{1{-}\tau_{\mathrm{ws}}(E)} & -\sqrt{\tau_{\mathrm{ws}}(E)[1{-}\tau_\mathrm{int}(E)]} & -\sqrt{\tau_\mathrm{ws}(E)\tau_\mathrm{int}(E)[1{-}\tau_\mathrm{res}(E)]} & \sqrt{\tau_\mathrm{int}(E)\tau_\mathrm{res}(E)\tau_\mathrm{ws}(E)} \\
        \sqrt{\tau_\mathrm{ws}(E)} & -\sqrt{[1{-}\tau_\mathrm{int}(E)][1{-}\tau_\mathrm{ws}(E)]} & -\sqrt{\tau_\mathrm{int}(E)[1{-}\tau_\mathrm{ws}(E)][1{-}\tau_\mathrm{res}(E)]} & \sqrt{\tau_\mathrm{int}(E)\tau_\mathrm{res}(E)[1{-}\tau_\mathrm{ws}(E)]} \\
        0 & \sqrt{\tau_\mathrm{int}(E)} & -\sqrt{[1{-}\tau_\mathrm{res}(E)][1{-}\tau_\mathrm{int}(E)]} & \sqrt{\tau_\mathrm{res}(E)[1{-}\tau_\mathrm{int}(E)]} \\
        0 & 0 & \sqrt{\tau_\mathrm{res}(E)} & \sqrt{1{-}\tau_\mathrm{res}(E)}
    \end{pmatrix},\quad\,\\
    J_\alpha^\Xi=\frac{1}{h}\int \dl E\,x^\Xi_\alpha\sum_{\beta}|\mathscr{S}_{\alpha\beta}|^2(f_\alpha-f_\beta),\\
    \begin{split}
        S_{J_\alpha^\Xi J_\beta^\Xi}=&\frac{2}{h}\int\dl E\,x^\Xi_\alpha x^\Xi_\beta\left\{\delta_{\alpha\beta}\sum_\gamma|\mathscr{S}_{\alpha\gamma}|^2 (F_{\alpha\gamma}+F_{\gamma\alpha})-F_{\alpha\alpha}|\mathscr{S}_{\beta\alpha}|^2-F_{\beta\beta}|\mathscr{S}_{\alpha\beta}|^2\right\}\\
        &-\frac{2}{h}\int\dl E\,x^\Xi_\alpha x^\Xi_\beta\mathrm{Re}\left\{\left[\sum_\gamma(f_\alpha-f_\gamma)\mathscr{S}_{\alpha\gamma}^*\mathscr{S}_{\beta\gamma}\right]\left[\sum_\gamma(f_\beta-f_\gamma)\mathscr{S}_{\alpha\gamma}\mathscr{S}_{\beta\gamma}^*\right]\right\}, 
    \end{split}
\end{gather}}where we have defined $x^N_\alpha = 1$, $x^Q_\alpha=E-\mu_\alpha$, and $F_{\alpha\beta} = f_\alpha(1 - f_\beta)$. Note that the scattering matrix is written in the $\{\L,\R,2,1\}$ basis and the energy dependence of the functions has been omitted in the integrands.
\end{widetext}

\bibliography{biblio.bib}

\end{document}


\title{Supplementary figures for: Precision of an autonomous demon exploiting nonthermal resources and information}

\author{Juliette Monsel}
\affiliation{Department of Microtechnology and Nanoscience (MC2), Chalmers University of Technology, S-412 96 G\"oteborg, Sweden\looseness=-1}
\author{Matteo Acciai}
\affiliation{The Abdus Salam International Center for Theoretical Physics, Strada Costiera 11, 34151 Trieste, Italy}
\affiliation{Scuola Internazionale Superiore di Studi Avanzati, Via Bonomea 256, 34136, Trieste, Italy}
\affiliation{Department of Microtechnology and Nanoscience (MC2), Chalmers University of Technology, S-412 96 G\"oteborg, Sweden\looseness=-1}
\author{Didrik Palmqvist}
\affiliation{Department of Microtechnology and Nanoscience (MC2), Chalmers University of Technology, S-412 96 G\"oteborg, Sweden\looseness=-1}
\author{Nicolas Chiabrando}
\affiliation{Department of Microtechnology and Nanoscience (MC2), Chalmers University of Technology, S-412 96 G\"oteborg, Sweden\looseness=-1}
\author{Rafael S\'anchez}
\affiliation{Departamento de F\'isica Te\'orica de la Materia Condensada, Universidad Aut\'onoma de Madrid, 28049 Madrid, Spain\looseness=-1}
\affiliation{Condensed Matter Physics Center (IFIMAC), Universidad Aut\'onoma de Madrid, 28049 Madrid, Spain\looseness=-1}
\affiliation{Instituto Nicol\'as Cabrera (INC), Universidad Aut\'onoma de Madrid, 28049 Madrid, Spain\looseness=-1}
\author{Janine Splettstoesser}
\affiliation{Department of Microtechnology and Nanoscience (MC2), Chalmers University of Technology, S-412 96 G\"oteborg, Sweden\looseness=-1}

\date{\today}

\maketitle
\vspace{-0.3cm}

In this Supplemental Material, we analyze in more details the performances of the refrigerator, first, by looking at various currents, noises and, cross-correlations appearing in the quantities plotted in the main text (in particular in the correlation coefficients from Fig.~7) (Sec.~\ref{SM:currents and noises}), second by analyzing the correlation coefficient between $J^I$, $\Jin$ and $J^I$, $\Jtrans$ (Sec.~\ref{app:extra figures:r}), then by studying the efficiency (Sec.~\ref{app:extra figures:efficiency}) and, finally, by examining the performance quantifiers associated to several uncertainty relations, including the so-called Clock Uncertainty Relation (CUR) \cite{Prech2025Sep} which is not discussed in the main text (Sec.~\ref{app:extra figures:URs}).
We also take a closer look at the noise approximation discussed in Appendix C.3 in the main text in Sec.~\ref{SM:approx noise}, quantifying the contributions of the time correlations to the noises and plotting the Pearson correlation coefficient as defined in Ref.~\cite{Freitas2021Mar}. In Sec.~\ref{SM:cycles}, we study the stochastic cycle rates, probabilities and durations. Finally, in Sec.~\ref{SM:cooling extra}, we show that, in regime (II), refrigeration is still possible when the reservoir that is being cooled down is colder than the coldest resource reservoir, namely when $T_\R < T_\C$.

\section{Average currents, noises and cross-correlations}\label{SM:currents and noises}

\begin{figure}[h]
    \includegraphics[width=0.85\linewidth]{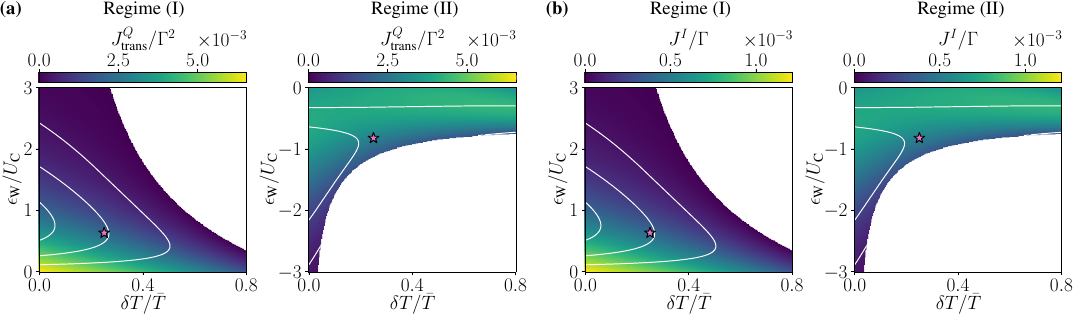}\vspace{-0.2cm}
        \caption{\label{fig:Jtrans and JI}
           Average (a)~transverse current $\Jtrans$ and (b)~information current $J^I$ as functions of $\eW$ and $\dT = T_\L - T_\R$. The points corresponding to the exact parameters of scenarios (I) and (II), see Table~I in the main text, are indicated by a purple star in the plots. The solid white lines indicate isolines of $\Pcool$ in all the plots (see Fig.~3(a) in the main text).}
\end{figure}

To complement the plots of the cooling power from the main text, we show here two other average currents of interest: the average transverse current $\Jtrans$ in Fig.~\subfigref{fig:Jtrans and JI}{a} and the information current $J^I$  in Fig.~\subfigref{fig:Jtrans and JI}{b}. Both behave in a very similar way as functions of the temperature difference in the working substance, $\dT$, and the energy level of dot W, $\eW$, but have different units and orders of magnitude.

\begin{figure}[h]
    \includegraphics[width=\linewidth]{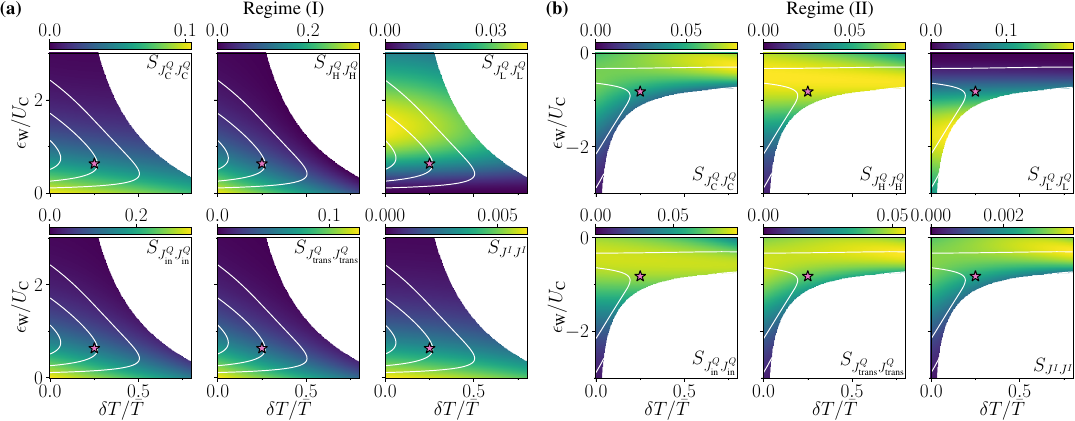}\vspace{-0.2cm}
        \caption{\label{fig:fluctuations}
            Noises (in units of $\Gamma^3$, except for $S_{J^I J^I}$ which is in units of $\Gamma$) for various currents as functions of $\eW$ and $\dT = T_\L - T_\R$ in (a)~regime~(I) and (b)~regime~(II). The points corresponding to the exact parameters of scenarios~(I) and (II), see Table~I in the main text, are indicated by a purple star in the plots. The solid white lines indicate isolines of $\Pcool$ in all the plots (see Fig.~3(a) in the main text).}
\end{figure}

Similarly, in Fig.~\ref{fig:fluctuations}, we plot the noises in the input heat current, $\Sin$, in the transverse heat current, $S_{\Jtrans\Jtrans}$, and in the information current, $S_{J^I J^I}$, as functions of $\eW$ and $\dT = T_\L - T_\R$. In particular, Fig.~\subfigref{fig:fluctuations}{a} further confirms that even if the average input heat current, $\Jin$, is zero, there are nonzero fluctuations which play a crucial role in the operation of the refrigerator, as explained in the main text and in Ref.~\cite{Acciai2024Feb}. Comparing Figs.~\subfigref{fig:fluctuations}{a} and \subfigref{fig:fluctuations}{b}, we see that the similarities between the transverse heat current and the information current go beyond the average currents and can also be seen in the noise.

\begin{figure}[h]
    \includegraphics[width=\linewidth]{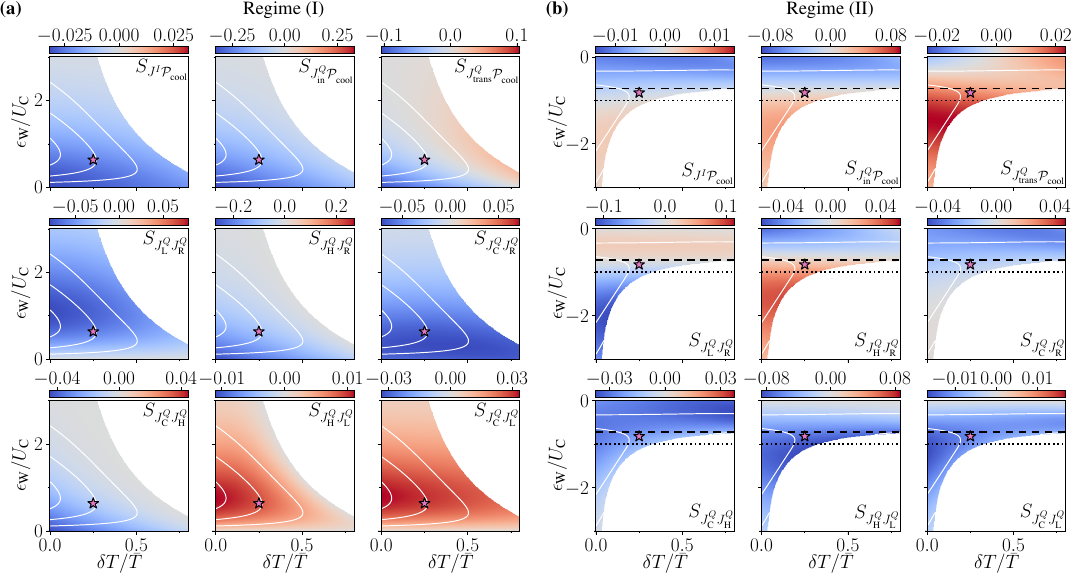}\vspace{-0.2cm}
    \caption{\label{fig:cross-corr}
        Cross-correlations (in units of $\Gamma^3$, except for those involving $J^I$ which are in units of $\Gamma^2$) between different pairs of currents as functions of $\eW$ and $\dT = T_\L - T_\R$ in (a)~regime~(I) and (b)~regime~(II). The points corresponding to the exact parameters of scenarios~(I) and (II), see Table~I in the main text, are indicated by a purple star in the plots. The solid white lines indicate isolines of $\Pcool$ in all the plots (see Fig.~3(a) in the main text). The horizontal dashed black line indicates $\eW = -\UH$ while the dotted black line indicates  $\eW = -\UC$.}
\end{figure}

Finally, to complement the Pearson correlation coefficients plotted in the main text (Fig.~7), we show here the non-normalized cross-correlations between different pairs of currents in Fig.~\ref{fig:cross-corr}. Unlike previously, comparing $S_{\Jtrans\Pcool}$ and $S_{J^I\Pcool}$ shows that $\Jtrans$ and $J^I$ do not correlate in the same way with the cooling power, even if they looked quite similar on average and at the noise level. In regime (II), in Fig.~\subfigref{fig:cross-corr}{b}, we also see that $S_{\JL\JR}$ and $S_{\Jin\Pcool}$ change sign (for a given $\dT$) when $Q_\R(\CH)$ changes sign, namely at $\eW = -\UH$ (dashed black line), highlighting the importance of the hot resource reservoir in this behavior. However, the sign change of $S_{J^I\Pcool}$ occurs when $Q_\R(\CC)$ changes sign, namely at $\eW = -\UC$ (dotted black line), which is in agreement with our previous finding that only the cooling process involving the cold resource reservoir exploits information as a resource.

\section{Efficiency of the refrigerator}\label{app:extra figures:efficiency}

The global efficiency of the device operating as a refrigerator can be defined as the ratio between entropy production rates in the reservoirs of the working substance and in the resource reservoirs \cite{InfoTrajPaper},
\begin{equation}\label{etaG}
    \eta_\text{global} = \frac{\beta_\L \JL + \beta_\R \JR}{-\beta_\C \JC - \beta_\H \JH}\ .
\end{equation}
Alternatively, we can also consider the information flowing from the working substance to the resource region as the relevant resource, giving rise to an information efficiency \cite{InfoTrajPaper}
\begin{equation}\label{etaI}
    \eta_\text{info} = \frac{\beta_\L \JL + \beta_\R \JR}{J^I}\ .
\end{equation}

\begin{figure}[h]
        \begin{minipage}[c]{0.45\linewidth}
            \includegraphics[width=\linewidth]{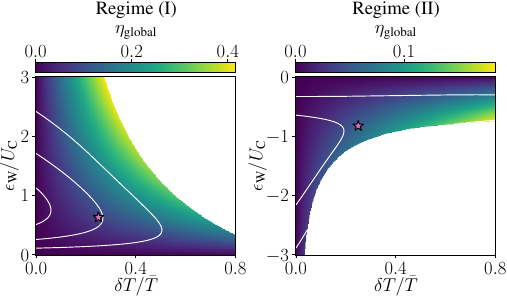}
        \end{minipage}\quad\quad
    \begin{minipage}{0.49\linewidth}
        \caption{\label{fig:efficiency}
            Efficiency $\eta_\text{global}$ [Eq.~\eqref{etaG}] as a function of $\eW$ and $\dT = T_\L - T_\R$. The points corresponding to the exact parameters of scenarios~(I) and (II), see Table~I in the main text, are indicated by a purple star in the plots. The solid white lines indicate isolines of $\Pcool$ in all the plots (see Fig.~3(a) in the main text).
        }
    \end{minipage}
    \vspace{-0.5cm}
\end{figure}

Even though $\eta_\text{info} \ge \eta_\text{global}$ in general, for the parameters considered here, $\eta_\text{info} \simeq \eta_\text{global}$, like in Ref.~\cite{InfoTrajPaper}. We have therefore only plotted $\eta_\text{global}$ as a function of $\eW$ and $\dT$ in Fig.~\ref{fig:efficiency}. We notice that the efficiency is overall larger in regime (I) than in regime (II) but only by a factor two approximately, unlike the precision which is orders of magnitude better in regime (II). Interestingly, $\eta_\text{global}$ increases with $\dT$ which is somewhat counterintuitive since the efficiency of a standard refrigerator tends to decrease with $\dT$ (since the associated Carnot efficiency is $T_\R/\dT = \bar{T}/\dT - 1/2$). At the demon condition, $\Jin = 0 = \JL + \JR = -\JH - \JC$, such that the efficiency becomes $\eta_\text{global} = \frac{(\beta_\R - \beta_\L) \Pcool }{(\beta_\H-\beta_\C) \JH}$ and for a fixed $\eW$, $\Pcool/\JH$ is approximately constant when changing $\dT$ while $\beta_\R - \beta_\L$ clearly increases.

\vspace{-0.2cm}

\section{Uncertainty relations}\label{app:extra figures:URs}

\begin{figure}[b]
    \includegraphics[width=\linewidth]{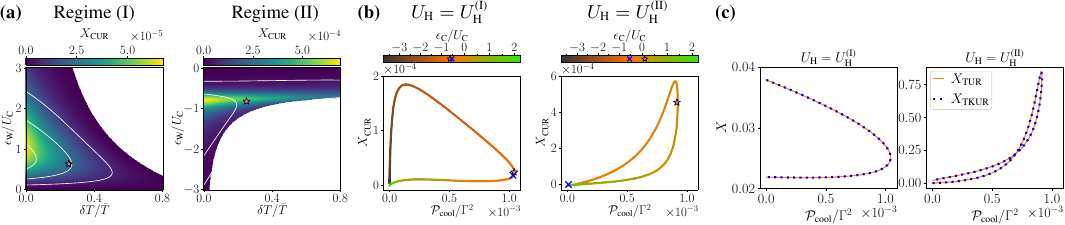}\vspace{-0.2cm}

    \caption{\label{fig:CUR}\label{fig:TKUR}
        (a)~$X_\text{CUR}$ as a function of $\eW$ and $\dT = T_\L - T_\R$. The points corresponding to the exact parameters of scenarios (I) and (II), see Table~I in the main text, are indicated by a purple star in the plots. The solid white lines indicate isolines of $\Pcool$ in all the plots (see Fig.~3(a) in the main text).
        (b)~Lasso plot for the $X_\text{CUR}$ obtained by varying $\eC$, like in Fig.~5 in the main text for the other performance quantifiers.
        (c)~Lasso plots of $X_\text{TUR}$ (solid orange) and $X_\text{TKUR}$ (dashed blue) obtained by varying $\eC$, like in Fig.~5 in the main text.
    }
\end{figure}

\begin{figure}[h]
    \begin{minipage}[t]{0.52\linewidth}
        \includegraphics[width=0.85\linewidth]{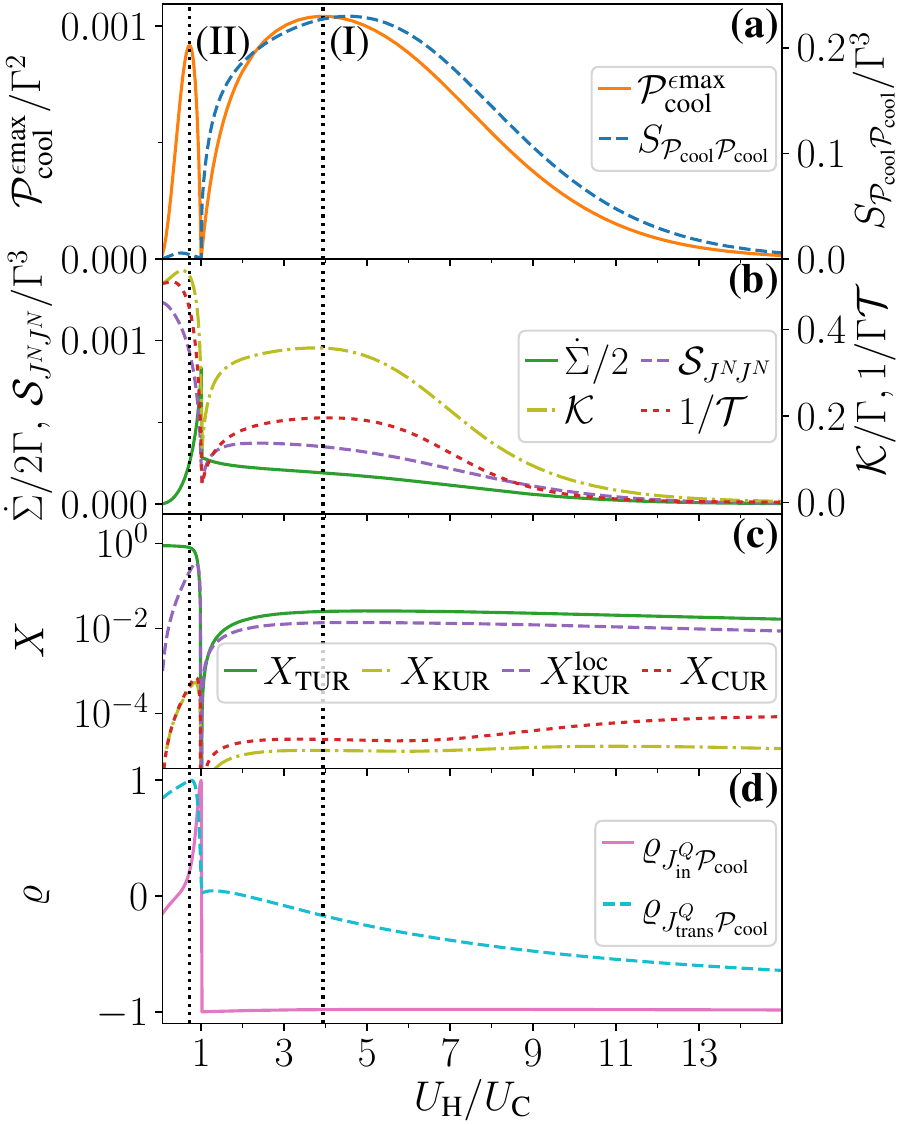}\vspace{-0.2cm}
        \caption{\label{fig:Q analysis at max power}
            (a)~Average cooling power and cooling power fluctuations, (b)~quantities in the denominator of the various uncertainty relation quantifiers $Q$: entropy production, activity, noise in the particle current, and inverse mean residual time, (c)~uncertainty relation quantifiers~$Q$ as defined in Sec.~II.B of the main text and in Eq.~\eqref{Q_CUR}, (d)~correlation coefficients between $\Pcool$ and, respectively, $\Jin$ and $\Jtrans$.
            As in Fig.~2, all quantities are plotted as functions of the interaction strength $\UH$, for $\eW$ and $\eC$ maximizing the cooling power and with $\eH$ chosen such that $\Jin = 0$.
        }

    \end{minipage}\hfill    \begin{minipage}[t]{0.45\linewidth}
        \includegraphics[width=\linewidth]{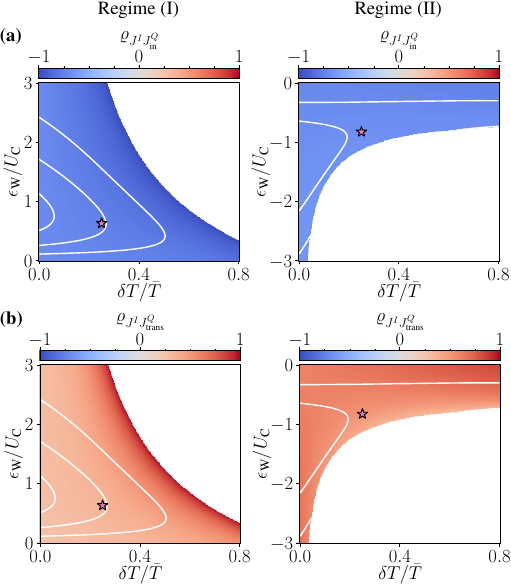}\vspace{-0.2cm}
        \caption{\label{fig:r extra}
            Pearson correlation coefficient from the FCS [Eq.~(9) in the main text] for (a)~$J^I$ and $\Jin$ and (b)~$J^I$ and $\Jtrans$ as functions of $\eW$ and $\dT = T_\L - T_\R$. The points corresponding to the exact parameters of scenarios~(I) and (II), see Table~I in the main text, are indicated by a purple star in the plots. The solid white lines indicate isolines of $\Pcool$ in all the plots (see Fig.~3(a) in the main text).
        }
    \end{minipage}
\end{figure}

In Ref.~\cite{Prech2025Sep}, a tighter bound on the precision than the KUR featuring the total activity is established, dubbed the Clock Uncertainty Relation (CUR). It is based on the mean residual time $\mathcal{T}$---that is the interval expected before the first jump is observed in a sequence, when observations start from an arbitrary time.
This translates into the performance quantifier
\begin{equation}\label{Q_CUR}
    X_\text{CUR} = \frac{\Pcool^2}{\Scool}\mathcal{T} \le 1\ ,
\end{equation}
with $\mathcal{T} = \sum_{hwc}\frac{\p_{hmc}}{\Gamma_{hwc}}$ and $\Gamma_{hwc} = -[W(0)]^{hwc}_{hwc}$. The CUR is tighter than the KUR since $\K \le \mathcal{T}^{-1}$ \cite{Prech2025Sep}, namely $X_\text{CUR} \ge X_\KUR$. However, looking at $X_\text{CUR}$ as function of $\eW$ and $\dT$ in Fig.~\subfigref{fig:CUR}{a}, we see that it behaves like $Q_\KUR$ in Fig.~4(b) in the main text. In regime (I), $X_\text{CUR}$ is about two times as large as $X_\KUR$, the order of magnitude thus remains $10^{-5}$, and, in regime (II), $X_\text{CUR}\simeq X_\KUR$. We make similar observations in Fig.~\subfigref{fig:Q analysis at max power}{c} for $X_\text{CUR}$ as a function of $\UH$. We are, however, optimizing our device for cooling power, while saturating the CUR is relevant, as indicated by the name, for maximizing clock precision \cite{Prech2025Sep}.
Finally, comparing the lasso plots for the CUR, Fig.~\subfigref{fig:CUR}{b}, and the KUR, Fig.~5(c) in the main text, we see again that they are quite similar, with slightly higher maximum values for $X_\text{CUR}$ but a somewhat worse trade-off between $X_\text{CUR}$ and the cooling power.

To get more insights in the performance quantifiers, we have plotted in Fig.~\ref{fig:Q analysis at max power} all the quantities involved in the $X$s, namely the trade-off based performance quantifiers, as functions of $\UH$ for the parameters maximizing the cooling power (that is like in Fig.~2 in the main text), as well as the $X$s themselves. The activity $\K$, the inverse mean residual time $1/\mathcal{T}$ [both on the right axis in Fig.~\subfigref{fig:Q analysis at max power}{b}], and the particle current noise $S_{J^NJ^N}$ (left axis) have similar behavior but different values, such that $X_\KUR$, $X_\text{CUR}$, and $X_\KUR^\text{loc}$ behave similarly while $X_\KUR^\text{loc}$ is orders of magnitude larger. The entropy production rate $\dot{\Sigma}$ has also a similar behavior at large $\UH \gg \UC$, but is very different for $\UH < \UC$, resulting in a plateau of $X_\TUR$ at almost 1 ($X_\TUR \simeq 0.89$ almost up to $\UH = \UC$). \\

We also looked at the TKUR \cite{Vo2022Sep}, unifying the KUR and TUR bounds in a more complicated quantity involving both the entropy production rate and the activity. We found that in the cases considered here, the TKUR bound is almost equal to the TUR one, as shown in Fig.~\ref{fig:TKUR}, which therefore appears to be the most appropriate one to evaluate the performance of our refrigerator.

\section{Correlation coefficients}\label{app:extra figures:r}

In Fig.~\ref{fig:r extra}, we see that in both regime (I) and (II), the information flow is anticorrelated with the input heat current $\Jin$ and correlated with the transverse heat current $\Jtrans$. This is because $\Jin = \JC + \JH$ while $\Jtrans = \JH - \JC$ and $J^I$ is strongly correlated with $\JC$ and only weakly correlated with $\JH$. This is consistent with the finding from Ref.~\cite{InfoTrajPaper} that the information-powered cooling (Maxwell demon-like) only involves the cold resource reservoir. The anticorrelation with $\Jin$ is stronger in regime (I) than in regime (II), while it is the opposite for the correlation with $\Jtrans$. $J^I$ being a resource enabling the refrigeration, this is consistent with the findings from the analysis of Figs.~7(b) and 7(c) in the main text that the cooling power is strongly anticorrelated with $\Jin$ in regime (I), while it is strongly correlated with $\Jtrans$ in regime (II).

Additionally, in Fig.~\subfigref{fig:Q analysis at max power}{d}, we have plotted the correlation coefficients $\varrho_{\Jin\Pcool}$ and $\varrho_{\Jtrans\Pcool}$ to confirm the analysis from Fig.~7 in the main text, showing that, more generally, for $\UH < \UC$, the cooling power is correlating to $\Jtrans$ while it is conversely anticorrelating to $\Jin$ for $\UH > \UC$. This highlights again how different the operating principles of the refrigerator are in the regimes $\UH < \UC$ and $\UH > \UC$.

\section{Noise approximations in terms of trajectory quantities}\label{SM:approx noise}

\begin{figure*}[h]
    \includegraphics[width=\linewidth]{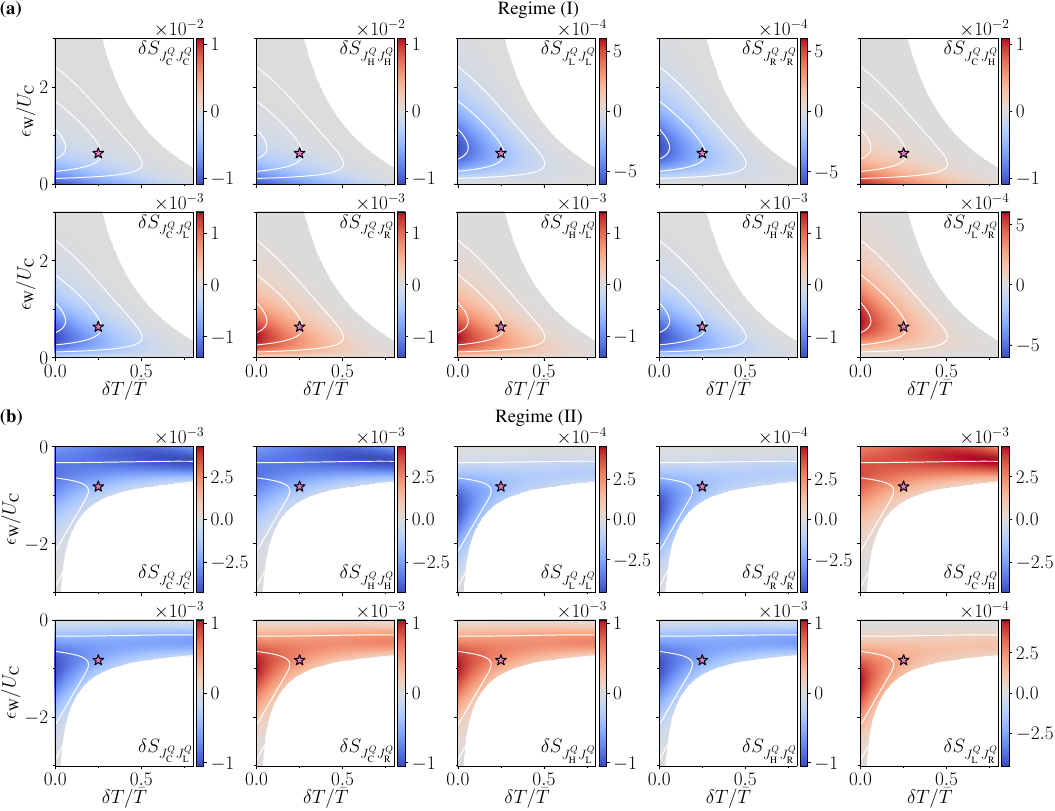}\vspace{-0.5cm}
    \caption{\label{fig:time corr}
        Contribution of time correlations (second, third and fourth terms in Eq.~(C8) in the main text) to the current noises and cross-correlations, $\delta S_{J_\alpha^QJ_\alpha^Q} =  S_{J_\alpha^QJ_\alpha^Q} - \text{var}(Q_\alpha)/\tcyc$ and  $\delta S_{J_\alpha^Q J_{\delta}^Q} =  S_{J_\alpha^QJ_{\delta}^Q} - \text{cov}(Q_\alpha, Q_{\delta})/\tcyc$,  as a function of $\eW$ and $\dT = T_\L - T_\R$. The points corresponding to the exact parameters of scenarios~(I) and (II), see Table~I in the main text, are indicated by a purple star in the plots. The solid white lines indicate isolines of $\Pcool$ in all the plots (see Fig.~3(a) in the main text). All quantities are in units of $\Gamma^3$.
    }
\end{figure*}

\subsection{Noise and cross-correlations}\label{SM:approx noise:dS}

To evaluate the precision of the noise approximation discussed in Appendix C.3, we plot the contribution of time correlations to the current noises and cross-correlations, $\delta S = S - S^\approx$, in Fig.~\ref{fig:time corr}. Comparing to the noises and cross-correlations in Figs.~\ref{fig:fluctuations} and \ref{fig:cross-corr}, we find that this term typically contributes well below $\sim$ 10\%, except in limited regions of cross-correlations involving one of the resource reservoir H or C, where it can reach 20\%.

\subsection{Pearson correlation coefficient}\label{SM:approx noise:Pearson}

\begin{figure}[tbh]
    \includegraphics[width=\linewidth]{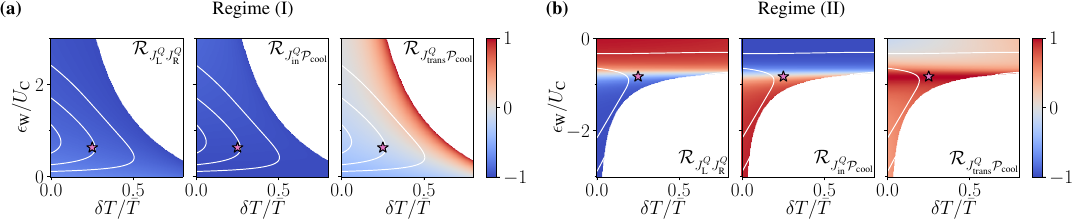}
    \caption{\label{fig:Pearson}
        Pearson correlation coefficient $\mathcal{R}$ as defined in Ref.~\cite{Freitas2021Mar} for different pairs of heat currents as functions of $\eW$ and $\dT = T_\L - T_\R$. The points corresponding to the exact parameters of scenarios (I) and (II), see Table~I in the main text, are indicated by a purple star in the plots. The solid white lines indicate isolines of $\Pcool$ in all the plots (see Fig.~3(a) in the main text).}
\end{figure}

As discussed in Appendix C.7 in the main text, the Pearson correlation coefficient defined in Ref.~\cite{Freitas2021Mar} differs from ours since it is defined from trajectory quantities rather than from the FCS quantities. However, this difference is expected to be very small in the cases we are interested in since we found that the noises and cross-correlations are well approximated by $S^\approx$, which corresponds to the noise/cross-correlation defined from trajectory quantities (see Sec.~\ref{SM:approx noise:dS}). This is confirmed by Fig.~\ref{fig:Pearson} which is virtually identical to panels (a)-(c) of Fig.~7 in the main text.

\section{Cycle duration and probabilities}\label{SM:cycles}

\begin{figure}[h]
    \includegraphics[width=\linewidth]{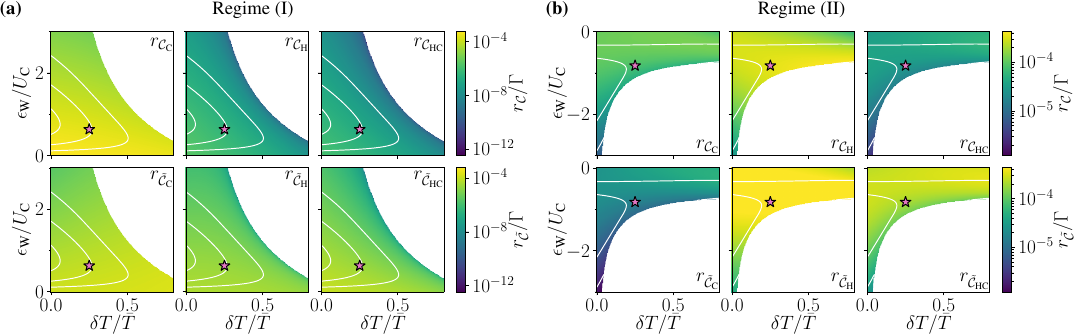}\vspace{-0.2cm}
    \caption{\label{fig:rates1}
         Cycle rates $r_\cC$  as a function of $\eW$ and $\dT = T_\L - T_\R$ in (a)~regime~(I) and (b)~regime~(II). The points corresponding to the exact parameters of scenarios~(I) and (II), see Table~I in the main text, are indicated by a purple star in the plots. The solid white lines indicate isolines of $\Pcool$ in all the plots (see Fig.~3(a) in the main text).
    }
\end{figure}

Figures \subfigref{fig:rates1}{a} and \subfigref{fig:rates1}{b} display the cycle rates $r_\cC$, in regimes (I) and (II) respectively, showing that the dominant process in regime (I) is the one involving only the cold resource reservoir and the dominant process in regime (II) is the one involving only the hot resource reservoir, in the findings for scenarios (I) and (II) in Ref.~\cite{InfoTrajPaper}.

\newpage
Figures \subfigref{fig:rates2}{a} and \subfigref{fig:rates2}{b} complement Fig.~16 in Appendix C.6 in the main text by showing the average duration of a cycle, $\tcyc$, the cycle probabilities, $\pi(\cC)$, and the cycle rates $\rC$ as functions of $\eW$ for a fixed $\dT$. We see that in regime (I), $\tcyc$ varies little and therefore $\pi(\cC)$ and $\rC$ have the same behavior when changing $\eW$. Conversely, in regime (II), $\tcyc$ decreases significantly when $\eW$ goes toward zero (due to the increased activity in the system) which makes the cycle rates increase even though the cycle probabilities are decreasing.\\

\begin{figure}[h]
    \includegraphics[width=0.95\linewidth]{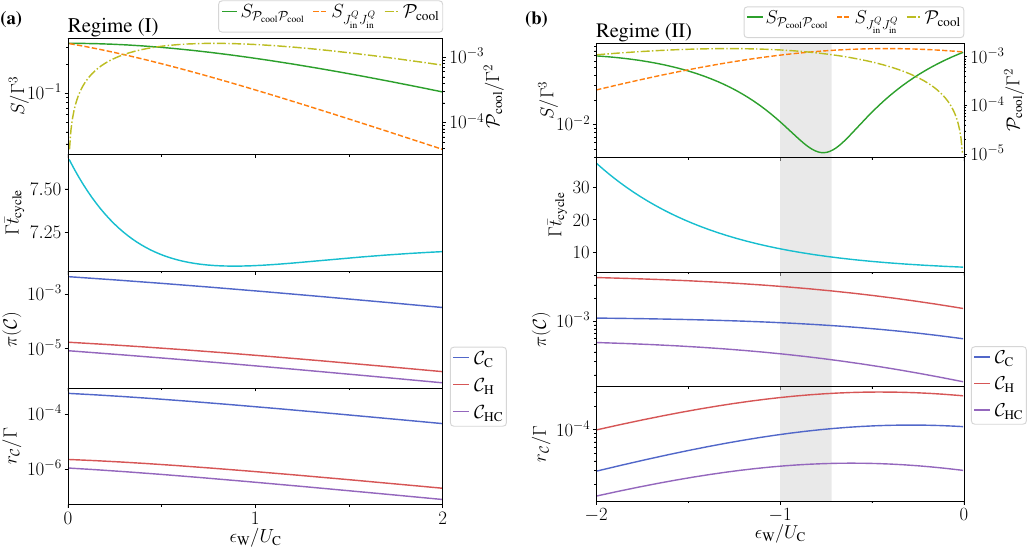}\vspace{-0.2cm}
    \caption{\label{fig:rates2}
        Noises, average cycle time $\tcyc$, cycle probabilities $\pi(\cC)$, and cycle rates $r_\cC$ as functions of $\eW$ for $\dT/\Gamma = 0.05$ in (a)~regime~(I) and (b)~regime~(II). Given that the entropy production $\Sigma(\cC)$ is approximately constant, $\pi(\bcC) = \e^{-\Sigma(\cC)}\pi(\cC)$ is just shifted compared to $\pi(\cC)$.  The gray-shaded area in panel (b) indicates the range of values of $\eW$ where $Q_\R(\CC)$ is positive while $Q_\R(\CH)$ is negative.
        }
\end{figure}

\newpage

\section{Cooling with $T_\R < T_\C$ in Regime (II)}\label{SM:cooling extra}
%

\begin{figure}[h]
    \includegraphics[width=0.95\linewidth]{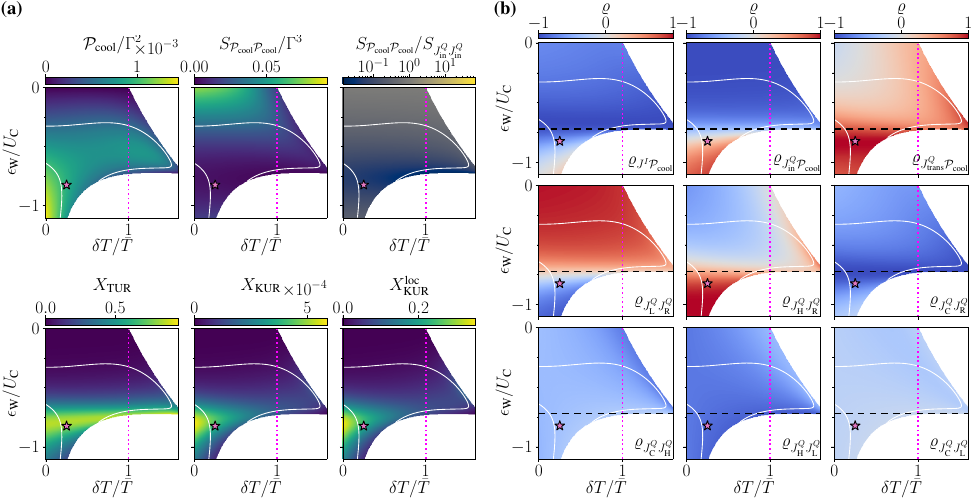}
    \caption{\label{fig:II 2}(a)~Various performance quantifiers of the refrigeration,  (b)~Pearson correlation coefficients in regime (II), see Table~I in the main text, for a larger range of temperature differences $\dT = T_\L - T_\R$ but a narrower range of values of $\eW$. The values of $\dT$ on the right of the vertical dotted purple line are such that $T_\R < T_\C$. In panel (b), the black dashed line indicates $\eW = -\UH$.}
\end{figure}

In the main text, we considered a range of temperature differences, $\dT = T_\L - T_\R$, such that the cold resource reservoir is always at the coldest temperature in the setup. However, this is not a necessary condition to achieve refrigeration. In this section, we extend the range of values of $\dT$ to beyond $\bar{T}$, meaning that $T_\R$ becomes lower than $T_\C$. In regime (I), the demon condition can no longer be fulfilled for $T_\R < T_\C$. By contrast, in regime (II), we find that we can still fulfill the demon condition for some values of $\eW$ and even cool down reservoir R in the working substance, despite it being colder than the coldest reservoir in the resource region, as shown in Fig.~\subfigref{fig:II 2}{a} (see the part of the plots on the right of the dotted purple vertical line). Figure ~\ref{fig:II 2} shows how the various quantities studied in the main text behave when $\dT$ goes beyond $\bar{T}$.

\bibliography{biblio.bib}